\documentclass[10pt,journal,compsoc]{IEEEtran}

\ifCLASSOPTIONcompsoc
  \usepackage[nocompress]{cite}
  \usepackage{cite}
\fi

\usepackage{mathptmx}
\usepackage{amsmath}
\usepackage{amssymb}
\usepackage{amsfonts}
\usepackage{graphicx}
\usepackage{times}
\usepackage{pifont}
\usepackage{comment}
\usepackage{algorithm}
\usepackage{algpseudocode}
\usepackage{gensymb}
\usepackage{wrapfig}
\usepackage{array}
\usepackage{soul}
\usepackage{cancel}
\usepackage{color}
\usepackage{slashbox}
\usepackage{multirow}
\usepackage{array}
\usepackage{makecell}
\usepackage{hyperref}
\usepackage{CJKutf8}
\usepackage{caption}
\usepackage{subfigure}
\usepackage{colortbl}
\usepackage{threeparttable}
\usepackage[table]{xcolor}
\usepackage{booktabs}
\usepackage{wasysym}

\usepackage{enumitem}
\setitemize{noitemsep,topsep=1pt,parsep=0pt,partopsep=1pt}

\makeatletter
\algrenewcommand\ALG@beginalgorithmic{\footnotesize}
\makeatother

\newcommand{\PreserveBackslash}[1]{\let\temp=\\#1\let\\=\temp}
\newcolumntype{C}[1]{>{\PreserveBackslash\centering}p{#1}}
\newcolumntype{R}[1]{>{\PreserveBackslash\raggedleft}p{#1}}
\newcolumntype{L}[1]{>{\PreserveBackslash\raggedright}p{#1}}

\usepackage[normalem]{ulem}

\newcommand{\revision}[1]{\leavevmode{\color{blue}{#1}}}
\newcommand{\remark}[1]{\textcolor[RGB]{200, 0, 0}{#1}}

\setstcolor{red}
\newcommand{\removed}[1]{\leavevmode{\color{red}{\st{#1}}}}

\def \cleanversion{} %
\ifx\cleanversion\undefined
\else
 \renewcommand{\remark}[1]{} %
 \renewcommand{\removed}[1]{} 
 \renewcommand{\revision}[1]{#1}
\fi

\usepackage{calc}
\newlength\myheight
\newlength\mydepth
\settototalheight\myheight{Xygp}
\begin{document}
\begin{CJK*}{UTF8}{gbsn}

\title{Datasets of Visualization for Machine Learning}

\author{Can~Liu,
        Ruike~Jiang,
        Shaocong Tan,
        Jiacheng Yu,
        Chaofan Yang,
        Hanning Shao,
        Xiaoru~Yuan%
\IEEEcompsocitemizethanks{
\IEEEcompsocthanksitem Can Liu, Ruike Jiang, Shaocong Tan, Jiacheng Yu, Chaofan Yang, Hanning Shao, and Xiaoru Yuan are with Key Laboratory of Machine Perception (Ministry of Education), School of Intelligence Science and Technology, Peking University. E-mail: \{can.liu, jiangrk, jiachengyu, chaofanyang, hanning.shao, xiaoru.yuan\}@pku.edu.cn.
\IEEEcompsocthanksitem Xiaoru Yuan is also with National Engineering Laboratory for Big Data Analysis and Application, Peking University.
\IEEEcompsocthanksitem Xiaoru Yuan is the corresponding author.}%
\thanks{Manuscript received November xx, 2023; revised November xx, 2023.}}

\markboth{TRANSACTIONS ON XXX,~Vol.~XX, No.~XX, November~2023}%
{Liu et al.: VisDataset}

\IEEEtitleabstractindextext{%
\begin{abstract}
Datasets of visualization play a crucial role in automating data-driven visualization pipelines, serving as the foundation for supervised model training and algorithm benchmarking. In this paper, we survey the literature on visualization datasets and provide a comprehensive overview of existing visualization datasets, including their data types, formats, supported tasks, and openness.
We propose a what-why-how model for visualization datasets, considering the content of the dataset (what), the supported tasks (why), and the dataset construction process (how). This model provides a clear understanding of the diversity and complexity of visualization datasets.
Additionally, we highlight the challenges faced by existing visualization datasets, including the lack of standardization in data types and formats and the limited availability of large-scale datasets. 
To address these challenges, we suggest future research directions.
\end{abstract}

}

\maketitle
\IEEEdisplaynontitleabstractindextext

\section{Introduction}

\revision{In recent years, there has been a significant trend towards data-driven and automated processes in visualization~\cite{ai4vis}.
Within these automated processes, machine learning models are employed, trained on related datasets specifically to perform various tasks related to visualization~\cite{wang2022ml4vis}.
Consequently, datasets tailored for visualization purposes have become a crucial infrastructure component in these automated processes.
Their importance and utility are progressively being recognized and emphasized.}
\removed{Datasets of visualizations have been used for tasks such as visual design space analysis, visualization retrieval, and user experiments for a long time.
However, recently the trend has shifted toward the automation of data-driven visualization pipelines, 
visualization data has become a kind of data format.
The automation trend resulted in an increased demand for datasets of visualizations.}
Compared to datasets in other domains, such as ImageNet~\cite{imagenet} and VQA~\cite{antol2015vqa}, datasets of visualizations have higher diversity in terms of data types, data formats, and tasks.
The data types include visualizations, visual components, underlying data, and additional information.
\revision{The underlying data in a visualization can take various forms such as tables, networks, or text, and the results of the visualization can be displayed in a diverse range of formats. Within the visualization itself, there are numerous visual elements, including but not limited to axes, legends, and titles. Beyond the core visual components, additional information often accompanies the visualization, like a set of questions and answers related to it, enhancing its informational value.}
A broad task space for visualization relies on datasets of various data formats, for example, constructing visualizations from data, reverse-engineering data from visualizations, and answering visualization-related questions.
To support these tasks, datasets of visualizations typically consist of multiple data types and serve as the foundation for supervised model training and benchmarking.
We survey the literature on datasets of visualizations, provide an overview of existing datasets of visualizations, summarize their data format, scale, supported tasks, and openness, and aim to emphasize the challenges and future research directions in datasets of visualizations.

\revision{In this survey, we define a dataset of visualization as a collection of data used for training or testing machine learning models in the visualization process.
Datasets of visualizations can be classified into a multi-layer scheme, including the content of it, the tasks supported, and how they are constructed.}
\removed{This classification takes into account data format, scale, supported tasks, and openness.}
An example of a dataset of visualization is VizNet~\cite{viznet}, which includes a wide range of visualization design examples, from simple bar charts to more complex visualizations. Its goal is to provide a comprehensive overview of the design space. Another example is VizCommender~\cite{vizcommender}, which is a dataset focused on retrieval. It contains a large collection of visualization designs, as well as user preferences and search queries, making it suitable for developing visualization recommendation systems.
In terms of data format, existing datasets of visualizations range from raster images to vector graphics and even text-based formats. For example, the VizML dataset~\cite{vizml} is a data-driven visualization construction dataset that focuses on converting data into visual representations.
It contains both data tables and the corresponding visualization results, allowing the development of models that can automatically generate visualizations from data.

\begin{figure*}[!ht]
  \centering
  \includegraphics[width=.9\textwidth]{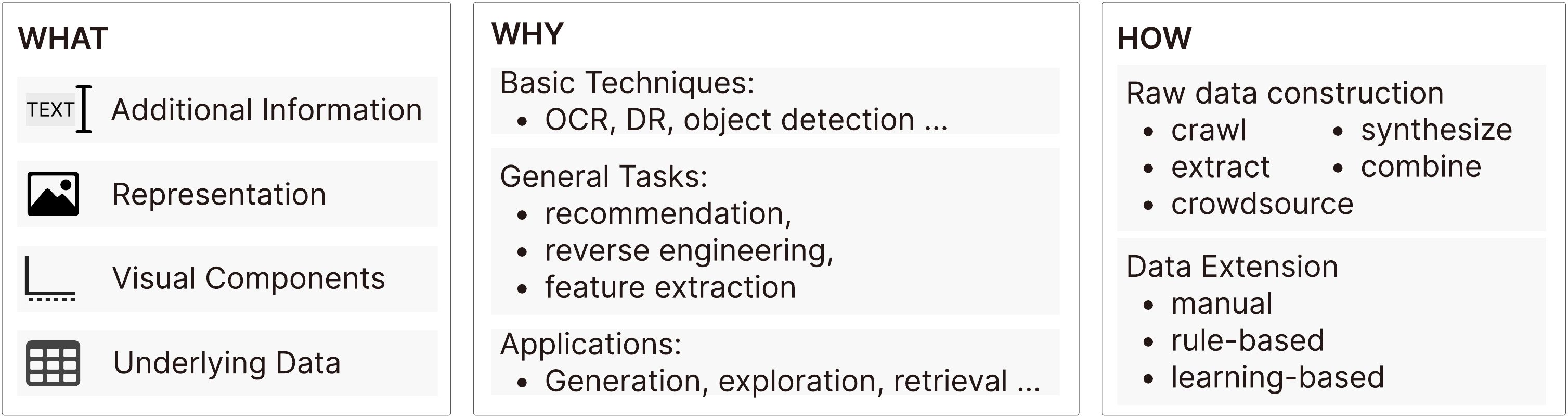}
   \caption{\label{fig:pipeline}
     The what-why-how model for the dataset of visualizations. What describes the content of the visualization, why describes the usage of the datasets, and how describes the construction methods of these datasets.}
\end{figure*}

Despite the consensus among researchers that constructing high-quality datasets of visualizations is crucial, the existing datasets still face numerous challenges. One of the major challenges is the lack of standardization in terms of data types and formats.
This lack of standardization makes it difficult to compare and integrate datasets from different sources, hindering the development of more advanced models and algorithms, especially in such an era when large datasets are needed.
Another challenge is the limited availability of large-scale datasets.
The current trend is towards automating data-driven visualization pipelines, which requires large-scale datasets for supervised model training.
To address these challenges, future research should focus on standardizing data types and formats, as well as increasing the openness of existing datasets.
Additionally, the intelligent construction of datasets of visualizations is also an important direction, aimed at building high-quality large-scale datasets with less manual labor.

We summarized the datasets of visualization presented in these papers using a what-why-how model, as illustrated in~\autoref{fig:pipeline}.

\begin{itemize}
    \item In terms of \textbf{what} is contained in the visualization dataset, we have decomposed the data type based on the formats and relationships present in the dataset. The visualization dataset may encompass underlying data, visualization components, presentation of the visualization, and accompanying information.
    \item As to \textbf{why} the visualization dataset is needed, it is established to support visualization tasks. We have categorized these tasks into three categories, namely basic techniques, general tasks, and applications.
    \item In regards to \textbf{how} the dataset is constructed, we have outlined several approaches to building a visualization dataset. The construction process has been summarized as comprising construction, synthesis, processing, annotation, and labeling.
\end{itemize}

\revision{``Why'' represents the machine learning tasks driven by the need for automation within the visualization process. In such tasks, it is common to employ a supervised dataset featuring correspondingly paired relationships to effectively train a model. ``What'' pertains to the specific nature of these tasks; for instance, a task like visualization recommendation necessitates a pair consisting of underlying data and its corresponding visualization. ``How'' dimension elucidates the methods employed in the collection of these datasets.}
The collection of datasets for visualization is shown in \autoref{tab:dataset}, categorized by what-why-how model.

        \begin{table*}[]
		\setlength{\abovecaptionskip}{0.1cm}
		\setlength{\belowcaptionskip}{-0.1cm}
		\caption{Summary of datasets of visualization for machine learning.}
		\label{tab:dataset}
		\centering
		\scriptsize
            \begin{threeparttable}
			\setlength{\tabcolsep}{0.5mm}
				\renewcommand\arraystretch{1}
				\rowcolors{2}{gray!20}{white}
				\begin{tabular}{l|cccccccc|cccccc|ccccc|ccc|cccccc|ccc|cccc|c|r|r|c}
    
    				\multicolumn{1}{c}{} & 
					  \multicolumn{8}{c}{\textbf{Data Types}} & 
					  \multicolumn{6}{c}{\textbf{\revision{Vis} Types}} & 
					  \multicolumn{8}{c}{\textbf{Data Construction}} &
					  \multicolumn{6}{c}{\textbf{Techniques}} & 
					  \multicolumn{3}{c}{\textbf{Tasks}} & 
					  \multicolumn{4}{c}{\textbf{\revision{User} Tasks}} &
                        	\multicolumn{1}{c}{} & 
                        	\multicolumn{1}{c}{} & 
                        	\multicolumn{1}{c}{}  \\
                         
					\textbf{Dataset} & 
					\rotatebox{90}{\textbf{VIS} }& 
					\rotatebox{90}{\textbf{NL} }& 
					\rotatebox{90}{\textbf{VIS Type} }& 
                    \rotatebox{90}{\textbf{Data} }& 
					\rotatebox{90}{\textbf{Mapping} }& 
					\rotatebox{90}{\textbf{VIS Element} }& 
					\rotatebox{90}{\textbf{Element Position} }& 
					\rotatebox{90}{\textbf{Others} }& 
					\rotatebox{90}{\textbf{Bar} }& 
					\rotatebox{90}{\textbf{Scatter} }& 
					\rotatebox{90}{\textbf{Line} }&
					\rotatebox{90}{\textbf{Area} }& 
					\rotatebox{90}{\textbf{Pie} }& 
					\rotatebox{90}{\textbf{Others} }& 
					\rotatebox{90}{\textbf{Crawling} }& 
					\rotatebox{90}{\textbf{Extraction} }& 
					\rotatebox{90}{\textbf{Crowdsourcing} }& 
					\rotatebox{90}{\textbf{Synthesizing} }& 
					\rotatebox{90}{\textbf{Combining} }&
					\rotatebox{90}{\textbf{Manual} }&  
					\rotatebox{90}{\textbf{Rule-based}}&
                    \rotatebox{90}{\textbf{Learning-based} }&  
                    \rotatebox{90}{\textbf{Object Detection} }&  
                    \rotatebox{90}{\textbf{Classification} }&  
                    \rotatebox{90}{\textbf{Regression} }&  
                    \rotatebox{90}{\textbf{Translation} }&  
                    \rotatebox{90}{\textbf{OCR} }&  
                    \rotatebox{90}{\textbf{DR} }&  
                    \rotatebox{90}{\textbf{Recommendation} }&  
                    \rotatebox{90}{\textbf{Reverse Engineering} }&  
                    \rotatebox{90}{\textbf{Feature Extraction} }&  
                    \rotatebox{90}{\textbf{Generation} }&  
                    \rotatebox{90}{\textbf{Retrieval } }&  
                    \rotatebox{90}{\textbf{Exploration } }&  
                    \rotatebox{90}{\textbf{Assessment} }&  
                    \rotatebox{90}{\textbf{Use for Learning} }&  
                    \rotatebox{90}{\textbf{Visualization size}}&  
                    \rotatebox{90}{\textbf{Data size}}&  
                    \rotatebox{90}{\textbf{Still Growing} }\\
					\midrule[1.5pt]

                    SNAP \cite{snapnets} &  &  &  & \CIRCLE &  &  &  &  &  &  &  &  &  &  & \CIRCLE &  &  &  &  &  &  &  &  &  &  &  &  &  &  &  &  &  &  &  &  &  & ~150 &  &  \\
ReVision \cite{revision} & \CIRCLE &  & \CIRCLE &  &  &  &  &  & \CIRCLE & \CIRCLE & \CIRCLE & \CIRCLE & \CIRCLE & \CIRCLE & \CIRCLE &  &  &  &  & \CIRCLE &  &  &  &  &  & \CIRCLE &  &  &  &  & \CIRCLE &  &  & \CIRCLE &  & \CIRCLE & ~2,500 &  &  \\
MASSVIS \cite{memorable} & \CIRCLE &  & \CIRCLE &  &  &  &  & \CIRCLE & \CIRCLE & \CIRCLE & \CIRCLE & \CIRCLE & \CIRCLE & \CIRCLE & \CIRCLE &  &  &  &  & \CIRCLE &  &  &  &  &  &  &  &  &  &  &  &  &  &  &  &  & 5,693 &  &  \\
DiagramFlyer \cite{chen2015diagramflyer} & \CIRCLE &  &  &  &  &  &  & \CIRCLE & \CIRCLE & \CIRCLE & \CIRCLE &  &  &  &  & \CIRCLE &  &  &  &  &  & \CIRCLE & \CIRCLE & \CIRCLE &  &  & \CIRCLE &  &  & \CIRCLE &  &  & \CIRCLE &  &  & \CIRCLE & ~300,000 &  &  \\
Aesthetics \cite{harrison2015infographic} & \CIRCLE &  &  &  &  &  &  &  &  &  &  &  &  & \CIRCLE & \CIRCLE &  &  &  &  & \CIRCLE &  &  &  &  &  &  &  &  &  &  &  &  &  &  &  &  & ~330 &  &  \\
WikiTable \cite{pasupat-liang-2015-compositional} &  & \CIRCLE &  & \CIRCLE &  &  &  &  &  &  &  &  &  &  & \CIRCLE &  &  &  &  & \CIRCLE &  &  &  &  &  &  &  &  &  &  &  &  &  &  &  &  &  & 2,108 D, 22,033 QA &  \\
Network Repository \cite{Rossi_2016_SIGKDD} & \CIRCLE &  &  & \CIRCLE &  &  &  &  &  &  &  &  &  & \CIRCLE &  &  &  &  & \CIRCLE &  & \CIRCLE &  &  &  &  &  &  &  &  &  &  &  &  &  &  &  & ~5,000 &  & \CIRCLE \\
Figureseer \cite{siegel2016figureseer} & \CIRCLE & \CIRCLE &  & \CIRCLE &  & \CIRCLE &  &  & \CIRCLE & \CIRCLE & \CIRCLE &  &  & \CIRCLE &  & \CIRCLE &  &  &  & \CIRCLE &  &  & \CIRCLE & \CIRCLE &  &  & \CIRCLE &  &  & \CIRCLE &  &  & \CIRCLE &  &  & \CIRCLE & ~60,000 &  &  \\
REV \cite{reverse} & \CIRCLE &  & \CIRCLE &  & \CIRCLE &  &  & \CIRCLE & \CIRCLE & \CIRCLE & \CIRCLE & \CIRCLE &  &  & \CIRCLE &  &  & \CIRCLE &  &  &  & \CIRCLE & \CIRCLE & \CIRCLE &  &  & \CIRCLE &  &  & \CIRCLE &  &  &  & \CIRCLE &  & \CIRCLE & 5,125 &  &  \\
ChartSense \cite{jung2017chartsense} & \CIRCLE &  & \CIRCLE &  &  &  &  &  & \CIRCLE & \CIRCLE & \CIRCLE & \CIRCLE & \CIRCLE & \CIRCLE & \CIRCLE &  &  &  &  & \CIRCLE & \CIRCLE &  &  & \CIRCLE &  &  &  &  &  &  & \CIRCLE &  & \CIRCLE &  &  & \CIRCLE & 6,997 &  &  \\
DataClips \cite{dataclip} & \CIRCLE &  & \CIRCLE & \CIRCLE &  &  &  & \CIRCLE & \CIRCLE & \CIRCLE &  & \CIRCLE & \CIRCLE & \CIRCLE & \CIRCLE &  &  &  &  & \CIRCLE &  &  &  &  &  &  &  &  &  &  &  &  &  &  &  &  & ~70 &  &  \\
BubbleView \cite{kim2017bubbleview} &  &  &  &  &  &  &  & \CIRCLE & \CIRCLE &  & \CIRCLE &  &  & \CIRCLE &  &  &  &  & \CIRCLE & \CIRCLE &  &  &  &  &  &  &  &  &  &  &  &  &  &  &  &  & 393 &  &  \\
FigureQA \cite{figureqa} & \CIRCLE & \CIRCLE &  &  &  &  &  & \CIRCLE & \CIRCLE & \CIRCLE & \CIRCLE &  &  &  &  &  &  & \CIRCLE &  &  & \CIRCLE &  &  &  &  & \CIRCLE &  &  &  &  & \CIRCLE & \CIRCLE &  &  &  & \CIRCLE & 100,000 & 1,000,000 QA &  \\
Sightline \cite{sightline} & \CIRCLE &  &  &  &  &  &  &  & \CIRCLE & \CIRCLE & \CIRCLE & \CIRCLE & \CIRCLE & \CIRCLE &  &  & \CIRCLE &  &  & \CIRCLE &  &  &  &  &  &  &  &  &  &  &  &  &  &  &  &  & 1,300 &  &  \\
Beagle \cite{beagle} & \CIRCLE &  & \CIRCLE &  &  &  &  &  & \CIRCLE & \CIRCLE & \CIRCLE & \CIRCLE & \CIRCLE & \CIRCLE & \CIRCLE &  &  &  &  &  &  & \CIRCLE &  &  &  & \CIRCLE &  &  &  &  & \CIRCLE &  &  & \CIRCLE &  & \CIRCLE & ~41,000 &  &  \\
ScatterNet \cite{ma2018scatternet} & \CIRCLE &  &  &  &  & \CIRCLE &  &  &  & \CIRCLE &  &  &  &  &  &  &  & \CIRCLE &  & \CIRCLE &  &  &  & \CIRCLE &  &  &  &  &  &  & \CIRCLE &  & \CIRCLE &  &  & \CIRCLE & 50,677 &  &  \\
DeepEye \cite{deepeye} & \CIRCLE &  &  &  &  &  &  & \CIRCLE & \CIRCLE & \CIRCLE & \CIRCLE & \CIRCLE & \CIRCLE & \CIRCLE &  &  &  & \CIRCLE &  & \CIRCLE &  &  &  & \CIRCLE &  &  &  &  & \CIRCLE &  &  & \CIRCLE &  &  &  & \CIRCLE & ~33,400 &  &  \\
DVQA \cite{Kafle2018} & \CIRCLE & \CIRCLE &  &  &  &  &  & \CIRCLE & \CIRCLE &  &  &  &  &  &  &  &  & \CIRCLE &  &  & \CIRCLE &  &  &  &  & \CIRCLE &  &  &  &  & \CIRCLE & \CIRCLE &  &  &  & \CIRCLE & 300,000 & 3,400,000 QA &  \\
VizByWiki \cite{lin2018vizbywiki} & \CIRCLE & \CIRCLE &  &  &  &  &  &  & \CIRCLE & \CIRCLE & \CIRCLE & \CIRCLE & \CIRCLE & \CIRCLE & \CIRCLE &  &  &  &  &  & \CIRCLE & \CIRCLE &  & \CIRCLE &  &  &  &  &  &  & \CIRCLE &  & \CIRCLE &  &  & \CIRCLE & ~3,000,000 &  & \CIRCLE \\
GMVR \cite{berger2018generative} &  &  &  & \CIRCLE &  &  &  &  &  &  &  &  &  & \CIRCLE &  &  &  & \CIRCLE &  &  & \CIRCLE &  &  &  & \CIRCLE &  &  &  &  &  & \CIRCLE & \CIRCLE &  &  &  & \CIRCLE & 200000 &  &  \\
Perception \cite{haehn2018evaluating} & \CIRCLE &  &  &  &  &  &  &  & \CIRCLE &  &  &  & \CIRCLE &  &  &  &  & \CIRCLE &  &  &  & \CIRCLE &  &  & \CIRCLE &  &  &  &  &  & \CIRCLE &  &  &  & \CIRCLE & \CIRCLE & 100,000 &  &  \\
Text-to-Viz \cite{text2viz} & \CIRCLE & \CIRCLE &  &  &  &  &  &  &  &  &  &  &  & \CIRCLE & \CIRCLE &  &  &  &  & \CIRCLE &  &  &  & \CIRCLE &  &  &  &  &  &  &  & \CIRCLE &  &  &  & \CIRCLE & 200 &  &  \\
D3 search \cite{d3search} & \CIRCLE &  & \CIRCLE & \CIRCLE &  &  &  & \CIRCLE & \CIRCLE & \CIRCLE & \CIRCLE & \CIRCLE & \CIRCLE & \CIRCLE & \CIRCLE &  &  &  &  &  & \CIRCLE &  &  &  &  &  &  &  &  &  &  &  &  &  &  &  & 7,860 &  &  \\
VizNet \cite{viznet} & \CIRCLE &  &  & \CIRCLE &  &  &  &  & \CIRCLE & \CIRCLE & \CIRCLE & \CIRCLE & \CIRCLE & \CIRCLE & \CIRCLE &  &  &  & \CIRCLE &  & \CIRCLE &  &  &  & \CIRCLE &  &  &  &  &  & \CIRCLE &  &  &  & \CIRCLE & \CIRCLE &  & ~31,000,000 &  \\
Timeline \cite{timeline} & \CIRCLE &  &  &  &  &  &  &  &  &  &  &  &  & \CIRCLE & \CIRCLE &  &  & \CIRCLE &  & \CIRCLE &  &  & \CIRCLE & \CIRCLE &  &  & \CIRCLE &  & \CIRCLE &  &  & \CIRCLE &  &  &  & \CIRCLE & 4,689 &  &  \\
VizML \cite{vizml} & \CIRCLE &  &  & \CIRCLE & \CIRCLE &  &  &  & \CIRCLE & \CIRCLE & \CIRCLE & \CIRCLE & \CIRCLE & \CIRCLE & \CIRCLE &  &  &  &  &  & \CIRCLE &  &  & \CIRCLE &  &  &  &  & \CIRCLE &  &  & \CIRCLE &  &  &  & \CIRCLE & ~1,000,000 &  &  \\
Data2Vis \cite{data2vis} & \CIRCLE &  &  & \CIRCLE & \CIRCLE &  &  &  & \CIRCLE & \CIRCLE & \CIRCLE & \CIRCLE & \CIRCLE & \CIRCLE &  &  &  & \CIRCLE &  &  & \CIRCLE &  &  &  &  & \CIRCLE &  &  & \CIRCLE &  &  & \CIRCLE &  &  &  & \CIRCLE & ~4,300 &  &  \\
Sherlock \cite{sherlock} &  &  &  & \CIRCLE &  &  &  &  &  &  &  &  &  &  &  &  &  &  & \CIRCLE &  & \CIRCLE & \CIRCLE &  &  &  &  &  &  &  &  &  &  &  &  &  & \CIRCLE &  & 686, 675 columns &  \\
InSituNet \cite{he2019insitunet} &  &  &  & \CIRCLE &  &  &  &  &  &  &  &  &  & \CIRCLE &  &  &  & \CIRCLE &  &  & \CIRCLE &  &  &  & \CIRCLE &  &  &  &  &  & \CIRCLE & \CIRCLE &  &  &  & \CIRCLE & 125,000 &  &  \\
DNN-VolVis \cite{hong2019dnn} &  &  &  & \CIRCLE &  &  &  &  &  &  &  &  &  & \CIRCLE &  &  &  & \CIRCLE &  &  & \CIRCLE &  &  &  & \CIRCLE &  &  &  &  &  & \CIRCLE & \CIRCLE &  &  &  & \CIRCLE & 10000 &  &  \\
AutoCaption \cite{liu2020autocaption} & \CIRCLE & \CIRCLE &  & \CIRCLE &  &  &  &  & \CIRCLE & \CIRCLE & \CIRCLE &  &  &  &  &  &  & \CIRCLE &  &  & \CIRCLE &  &  &  &  & \CIRCLE &  &  &  &  & \CIRCLE & \CIRCLE &  &  &  & \CIRCLE & 3,000 &  &  \\
Auto Annotation \cite{lai2020automatic} & \CIRCLE &  &  &  &  & \CIRCLE &  &  & \CIRCLE & \CIRCLE &  &  & \CIRCLE &  & \CIRCLE &  &  &  &  & \CIRCLE &  &  & \CIRCLE & \CIRCLE &  &  & \CIRCLE &  &  & \CIRCLE & \CIRCLE & \CIRCLE &  &  &  & \CIRCLE & ~400 &  &  \\
MultiViewLayout \cite{chen2020composition} & \CIRCLE &  &  &  &  &  &  & \CIRCLE &  &  &  &  &  &  &  & \CIRCLE &  &  &  & \CIRCLE &  &  &  &  &  &  &  &  &  &  &  &  &  &  &  &  & 360 &  &  \\
VisImages \cite{visimages} & \CIRCLE &  &  &  &  &  &  &  & \CIRCLE & \CIRCLE & \CIRCLE & \CIRCLE & \CIRCLE & \CIRCLE &  & \CIRCLE &  &  &  & \CIRCLE &  &  & \CIRCLE & \CIRCLE &  &  & \CIRCLE &  &  & \CIRCLE &  &  & \CIRCLE &  &  & \CIRCLE & 12,267 &  &  \\
InfoVIF \cite{infovif} & \CIRCLE &  &  & \CIRCLE &  & \CIRCLE & \CIRCLE & \CIRCLE &  &  &  &  &  & \CIRCLE & \CIRCLE &  &  &  &  & \CIRCLE &  & \CIRCLE & \CIRCLE & \CIRCLE &  &  & \CIRCLE &  &  & \CIRCLE &  &  &  & \CIRCLE &  & \CIRCLE & ~13,000 &  &  \\
Retrie-then-adapt \cite{qian2020retrieve} & \CIRCLE &  &  &  &  & \CIRCLE & \CIRCLE &  &  &  &  &  &  & \CIRCLE & \CIRCLE &  &  &  &  & \CIRCLE &  &  & \CIRCLE & \CIRCLE &  &  & \CIRCLE &  & \CIRCLE & \CIRCLE &  & \CIRCLE &  &  &  & \CIRCLE & 1,000 &  &  \\
InteractiveArticles \cite{hohman2020communicating} & \CIRCLE &  &  &  &  &  &  & \CIRCLE &  &  &  &  &  & \CIRCLE &  &  & \CIRCLE &  &  & \CIRCLE &  &  &  &  &  &  &  &  &  &  &  &  &  &  &  &  & ~60 &  &  \\
Leaf-QA \cite{chaudhry2020leaf} & \CIRCLE & \CIRCLE &  &  &  &  &  & \CIRCLE & \CIRCLE & \CIRCLE & \CIRCLE &  & \CIRCLE & \CIRCLE &  &  &  & \CIRCLE &  &  & \CIRCLE & \CIRCLE &  &  &  & \CIRCLE &  &  &  &  & \CIRCLE & \CIRCLE &  &  &  & \CIRCLE & 240,000 & 2,000,000 QA &  \\
Leaf-QA++ \cite{singh2020stl} & \CIRCLE & \CIRCLE &  &  &  &  &  & \CIRCLE & \CIRCLE & \CIRCLE & \CIRCLE &  & \CIRCLE & \CIRCLE &  &  &  & \CIRCLE &  & \CIRCLE & \CIRCLE &  &  &  &  & \CIRCLE &  &  &  &  & \CIRCLE & \CIRCLE &  &  &  & \CIRCLE & 244,000 & 2,500,000 QA &  \\
PlotQA \cite{methani2020plotqa} & \CIRCLE & \CIRCLE &  &  &  &  &  & \CIRCLE & \CIRCLE & \CIRCLE & \CIRCLE &  &  &  &  &  &  & \CIRCLE &  & \CIRCLE & \CIRCLE &  &  &  &  & \CIRCLE &  &  &  &  & \CIRCLE & \CIRCLE &  &  &  & \CIRCLE & 224,000 & 28,000,000 QA &  \\
Vis-QA \cite{Kim2020} & \CIRCLE & \CIRCLE &  &  &  &  &  & \CIRCLE & \CIRCLE &  & \CIRCLE &  &  &  & \CIRCLE &  &  &  &  & \CIRCLE &  &  &  &  &  &  &  &  &  &  &  &  &  &  &  &  & 52 & 629 QA &  \\
FigCAP \cite{Chen_2020_WACV} & \CIRCLE & \CIRCLE &  &  &  &  &  &  & \CIRCLE &  & \CIRCLE &  & \CIRCLE &  &  &  &  & \CIRCLE &  &  & \CIRCLE &  &  &  &  & \CIRCLE &  &  &  &  & \CIRCLE & \CIRCLE &  &  &  & \CIRCLE & 124,217 &  &  \\
VizCommender \cite{vizcommender} & \CIRCLE & \CIRCLE &  &  & \CIRCLE &  &  &  & \CIRCLE & \CIRCLE & \CIRCLE & \CIRCLE & \CIRCLE & \CIRCLE & \CIRCLE &  &  &  &  &  & \CIRCLE &  &  & \CIRCLE &  &  &  &  &  &  & \CIRCLE &  & \CIRCLE &  &  & \CIRCLE & 18,820 &  &  \\
ChartSeer \cite{Zhao2020a} & \CIRCLE &  &  & \CIRCLE & \CIRCLE &  &  &  & \CIRCLE & \CIRCLE & \CIRCLE & \CIRCLE &  &  &  &  &  & \CIRCLE & \CIRCLE &  &  &  &  &  &  &  &  & \CIRCLE &  &  & \CIRCLE &  &  & \CIRCLE &  & \CIRCLE & 9,925 &  &  \\
Calliope \cite{shi2020calliope} & \CIRCLE &  & \CIRCLE &  &  &  &  & \CIRCLE & \CIRCLE & \CIRCLE & \CIRCLE & \CIRCLE & \CIRCLE & \CIRCLE &  &  & \CIRCLE &  &  & \CIRCLE &  &  &  &  &  &  &  &  &  &  &  &  &  &  &  &  & 230 & 4,186 segments &  \\
DVAO \cite{engel2020deep} &  &  &  & \CIRCLE &  &  &  &  &  &  &  &  &  & \CIRCLE &  &  &  & \CIRCLE &  &  & \CIRCLE &  &  &  & \CIRCLE &  &  &  &  &  & \CIRCLE & \CIRCLE &  &  &  & \CIRCLE & 477 &  &  \\
CrisisVis \cite{covid19landscape} & \CIRCLE &  & \CIRCLE &  &  &  &  &  & \CIRCLE & \CIRCLE &  &  &  &  &  &  & \CIRCLE &  &  & \CIRCLE &  &  &  &  &  &  &  &  &  &  &  &  &  &  &  &  & 663 &  &  \\
ViralVis \cite{lee2021viral} &  &  &  &  &  &  &  &  & \CIRCLE & \CIRCLE & \CIRCLE & \CIRCLE &  & \CIRCLE & \CIRCLE &  &  &  &  & \CIRCLE &  &  &  &  &  & \CIRCLE &  &  &  &  & \CIRCLE &  &  & \CIRCLE &  & \CIRCLE & ~41,000 &  &  \\
VIS30K \cite{chen2021vis30k} & \CIRCLE &  &  &  &  &  &  & \CIRCLE &  &  &  &  &  & \CIRCLE &  & \CIRCLE &  &  &  & \CIRCLE &  &  & \CIRCLE & \CIRCLE &  &  & \CIRCLE &  &  & \CIRCLE &  &  & \CIRCLE &  &  & \CIRCLE & 29,689 &  &  \\
ExcelChart400K \cite{luo2021chartocr} & \CIRCLE &  & \CIRCLE & \CIRCLE &  &  &  &  & \CIRCLE &  & \CIRCLE &  & \CIRCLE &  & \CIRCLE &  &  &  &  &  &  & \CIRCLE &  &  &  &  & \CIRCLE &  &  & \CIRCLE &  &  & \CIRCLE &  &  & \CIRCLE & 386,966 &  &  \\
Visually29K \cite{Madan2021} & \CIRCLE &  &  &  &  & \CIRCLE & \CIRCLE &  &  &  &  &  &  & \CIRCLE & \CIRCLE &  &  &  &  & \CIRCLE &  &  & \CIRCLE & \CIRCLE &  &  & \CIRCLE &  &  & \CIRCLE &  &  & \CIRCLE & \CIRCLE &  & \CIRCLE & ~29,000 &  &  \\
AutoClips \cite{shi2021autoclips} & \CIRCLE &  &  & \CIRCLE &  &  &  & \CIRCLE &  &  &  &  &  & \CIRCLE & \CIRCLE &  &  &  &  & \CIRCLE &  &  &  &  &  &  &  &  &  &  &  &  &  &  &  &  & ~230 &  &  \\
Motion \cite{shi2021video} &  &  &  &  &  &  &  &  &  &  &  &  &  & \CIRCLE & \CIRCLE &  &  &  &  & \CIRCLE &  &  &  &  &  &  &  &  &  &  &  &  &  &  &  &  & 82 &  &  \\
Kineticharts \cite{lan2021kineticharts} &  &  &  & \CIRCLE &  &  &  & \CIRCLE &  &  &  &  &  & \CIRCLE & \CIRCLE &  &  &  &  & \CIRCLE &  &  &  &  &  &  &  &  &  &  &  &  &  &  &  &  & 259 &  &  \\
Data-GIF \cite{datagif} & \CIRCLE &  &  &  &  &  &  & \CIRCLE &  &  &  &  &  & \CIRCLE & \CIRCLE &  &  &  &  & \CIRCLE &  &  &  &  &  &  &  &  &  &  &  &  &  &  &  &  & 108 &  &  \\
NLV Corpus \cite{srinivasan2021collecting} & \CIRCLE & \CIRCLE &  &  &  &  &  &  &  &  &  &  &  & \CIRCLE &  &  &  & \CIRCLE &  & \CIRCLE &  &  &  & \CIRCLE &  &  &  &  &  &  & \CIRCLE & \CIRCLE &  &  &  & \CIRCLE &  & 893 Sentences &  \\
NL2VIS \cite{luo2021synthesizing} & \CIRCLE & \CIRCLE &  &  &  &  &  & \CIRCLE &  &  &  &  &  & \CIRCLE &  &  &  &  & \CIRCLE & \CIRCLE & \CIRCLE &  &  &  &  & \CIRCLE &  &  & \CIRCLE &  &  & \CIRCLE &  &  &  & \CIRCLE & 25,750 & 25,750 &  \\
SCICAP \cite{hsu-etal-2021-scicap-generating} & \CIRCLE & \CIRCLE & \CIRCLE &  &  &  &  &  &  &  & \CIRCLE &  &  &  & \CIRCLE & \CIRCLE &  &  &  &  &  & \CIRCLE &  &  &  & \CIRCLE & \CIRCLE &  &  &  & \CIRCLE & \CIRCLE &  &  &  & \CIRCLE & 358,972 & 358,972 &  \\
Table2Charts \cite{zhou2021table2charts} & \CIRCLE & \CIRCLE &  & \CIRCLE & \CIRCLE &  &  &  &  &  &  &  &  & \CIRCLE &  &  & \CIRCLE &  &  &  & \CIRCLE &  &  & \CIRCLE &  &  &  &  & \CIRCLE &  &  & \CIRCLE &  &  &  & \CIRCLE & 266,000 & 165,000 D &  \\
STNet \cite{han2021stnet} &  &  &  & \CIRCLE &  &  &  &  &  &  &  &  &  & \CIRCLE &  &  &  & \CIRCLE &  &  & \CIRCLE &  &  &  & \CIRCLE &  &  &  &  &  & \CIRCLE & \CIRCLE &  &  &  & \CIRCLE & 650 & 650 &  \\
ChartQA \cite{chartqa} & \CIRCLE & \CIRCLE &  &  &  &  &  & \CIRCLE &  &  &  &  &  & \CIRCLE & \CIRCLE &  &  &  &  & \CIRCLE &  & \CIRCLE &  &  &  & \CIRCLE &  &  &  &  & \CIRCLE & \CIRCLE &  &  &  & \CIRCLE &  & 9,600 Q, 23, 100 A &  \\
MultiVision \cite{Wu2022} & \CIRCLE &  &  &  &  &  &  &  & \CIRCLE & \CIRCLE & \CIRCLE &  &  & \CIRCLE &  &  &  &  & \CIRCLE &  &  &  &  & \CIRCLE &  &  &  &  & \CIRCLE &  &  & \CIRCLE &  &  &  & \CIRCLE &  & 3,920,941 QA &  \\
GoTreeScape \cite{gotreescape} & \CIRCLE &  &  & \CIRCLE & \CIRCLE &  &  &  &  &  &  &  &  & \CIRCLE &  &  &  & \CIRCLE &  &  &  &  &  &  &  &  &  & \CIRCLE &  &  & \CIRCLE &  &  & \CIRCLE &  & \CIRCLE & 62,340 & 62,340 &  \\
Pictorial \cite{shi2022supporting} &  &  &  &  &  &  &  &  & \CIRCLE &  & \CIRCLE & \CIRCLE & \CIRCLE & \CIRCLE & \CIRCLE &  &  &  & \CIRCLE & \CIRCLE &  &  &  &  &  &  &  &  &  &  &  &  &  &  &  &  & 1,371 & 1,371 &  \\
CoordNet \cite{han2022coordnet} &  &  &  & \CIRCLE &  &  &  &  &  &  &  &  &  & \CIRCLE &  &  &  & \CIRCLE &  &  & \CIRCLE &  &  &  & \CIRCLE &  &  &  &  &  & \CIRCLE & \CIRCLE &  &  &  & \CIRCLE & 32,000 & 32,000 &  \\
AutoTitle \cite{liu2023autotitle} &  & \CIRCLE &  & \CIRCLE &  &  &  &  &  &  &  &  &  &  &  &  &  & \CIRCLE & \CIRCLE & \CIRCLE &  & \CIRCLE &  &  &  & \CIRCLE &  &  & \CIRCLE &  &  & \CIRCLE &  &  &  & \CIRCLE &  & 6,000 fact-title &  \\
OldVisOnline \cite{zhang2023oldvis} & \CIRCLE &  & \CIRCLE &  &  &  &  &  & \CIRCLE & \CIRCLE & \CIRCLE & \CIRCLE & \CIRCLE & \CIRCLE & \CIRCLE &  &  &  &  & \CIRCLE &  & \CIRCLE &  & \CIRCLE &  &  &  &  &  &  &  &  &  & \CIRCLE &  & \CIRCLE & 13,511 &  & \CIRCLE \\
MoneyVis \cite{firat2023moneyvis} &  &  &  & \CIRCLE &  &  &  & \CIRCLE &  &  &  &  &  &  &  &  & \CIRCLE &  &  &  & \CIRCLE &  &  & \CIRCLE &  &  &  &  &  &  & \CIRCLE &  &  &  &  &  &  & 1 &  \\

                    \end{tabular}
                \begin{tablenotes}
				\scriptsize
				\item[*] Abbreviations for \textbf{\textit{Data Size}}: Data (D), Visualization (V), Question (Q), Answer (A).
        
                    \item[*] The online version can be accessed through: \url{https://visdataset.github.io/}.
			\end{tablenotes}
                \end{threeparttable}
	\end{table*}

\section{\revision{Related Surveys}}

\revision{Wang et al.~\cite{wang2022ml4vis} discussed machine learning models applied to visualization. Addressing artificial intelligence, Wu et al.~\cite{ai4vis} provided a review of AI techniques in VIS research. Davila et al.~\cite{davila2021chartmining} conducted a comprehensive survey of Chart Mining, compiling works that support the recovery of underlying data from charts.
Liu et al.~\cite{liu2023visualizationresources} made a detailed review of open collections of visualization resources, including papers that collect metadata, visualization survey papers that provide image browsing interfaces, visualization books, and websites containing examples of visualization programming libraries, etc.
Chen and Liu~\cite{chen2023state} proposed a survey focused on the creation of visualization corpora for automated chart analysis.
Differing from their approaches, our work primarily targets \textbf{datasets} as the research subject, analyzing the combination of data forms to support various machine learning tasks in visualization.}

\revision{
\section{Methodology}

\subsection{Definition and Scope}

A dataset of visualization is a collection of data used for training or testing machine learning models in the visualization process.
The following criteria must be met for a dataset to be considered in this paper:

\begin{itemize}
    \item The dataset should not consist solely of underlying data used for visualization purposes, such as tables (e.g., R datasets~\cite{rdataset}), networks, or volume data. For example, if a dataset only contains underlying data for scientific visualization, it is not within our scope. However, if a dataset includes derived batches of rendering results and supports deep learning work, it is considered within our purview.

    \item The dataset should primarily comprise data used in machine learning-related work within the visualization domain or data employed in the training or evaluation processes of machine learning models for visualization.

    \item The dataset should mainly encompass large-scale, annotated visualization data. This ensures that the dataset can facilitate the development and evaluation of machine learning models.

\end{itemize}

We collected works from IEEE VIS, EuroVis, and PacificVis that utilized machine learning models and employed training data in their methods. We traversed the datasets used in these works, as well as the works they cited and the works that cited them.

\subsection{Coding}

\autoref{fig:pipeline} provides the overall conceptual framework, categorized according to the "what-why-how" three-level structure.

From the "what" perspective, we discuss the components of a visualization dataset, including whether it contains underlying data, detailed components, its form of expression, and additional information.
We annotated each dataset with information on these dimensions.
Additionally, we also focus on the scale of the visualization dataset, including the number of visualizations and the quantity of other forms.
Regarding the tasks supported or potentially supported by the dataset (WHY), we annotate on three levels: basic techniques, visualization tasks, and user tasks.
As for how the dataset is constructed (HOW), we categorize the data into parts that can be directly scraped or synthesized, referred to as raw data, and parts that require additional annotation, termed data augmentation.

In the annotation process, all markings are done by at least two people. If there are inconsistencies, another person independently annotates before a discussion takes place. These details are all listed in \autoref{tab:dataset}.

On our supplementary website, visdataset.github.io, we provide an interactive system that supports filtering and selection based on several aspects, including tasks, data formats, and construction methods.
Details information also include links to papers and the corresponding datasets.

}

\begin{figure*}[!htb]
  \centering
  \includegraphics[width=\textwidth]{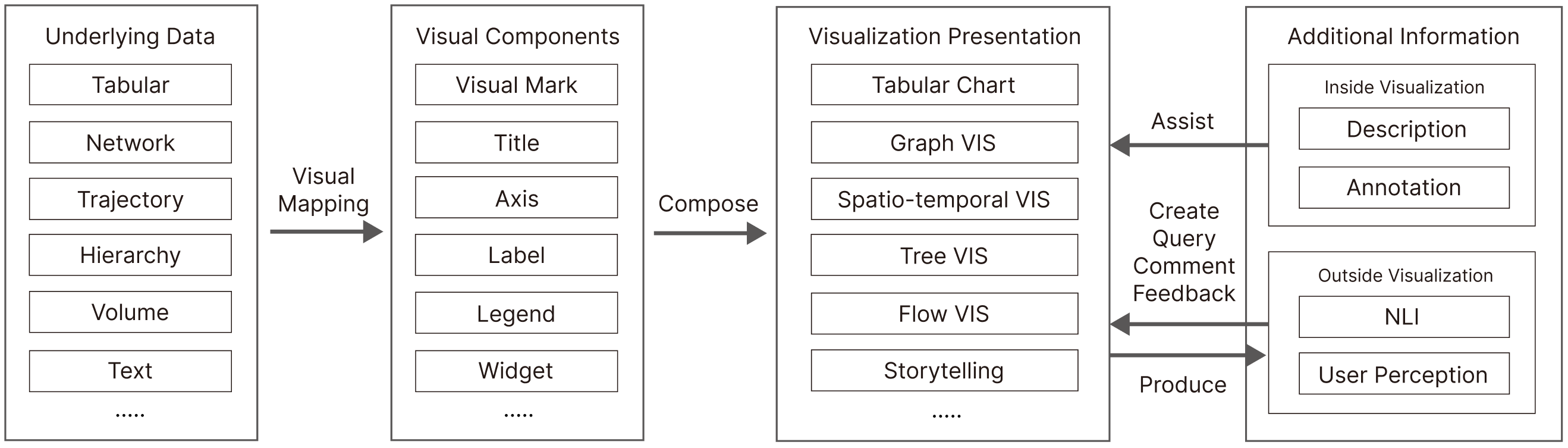}
   \caption{\label{fig:what}
     Relationship of different datasets in ``What". The underlying data are objects for visual mapping and rendering to build visualization components. These components are combined and laid out to compose a visualization presentation, which can be enhanced with NL, such as descriptions and annotations. NL can be an interface to create, query, comment, and give feedback on a visualization. In addition, different forms of visualization and their components can be used for tasks such as user perception experiments and machine learning training.}
\end{figure*}

\section{What: Content of Dataset}

\revision{Datasets of visualization are collections of data formats related to visualization, including underlying data, visualization results (components and presentation), and additional information, as depicted in~\autoref{fig:what}.
The underlying data refers to the content that needs to be visualized, such as tabular and network data.
The visualization results can be described through their internal components and presentation format.
Visualization components include the content of the visualization, encompassing visual elements that encode data and functional components like axes and legends. Visualization presentation refers to the type of visualization.
Additional information is also a key aspect of visualization.
While it is not a constitutive element of visualization, it enhances the expressiveness of the visualization, such as question and answer related to the visualization.}

\subsection{Underlying Data}

\begin{figure}[htb]
  \centering
  \includegraphics[width=\columnwidth]{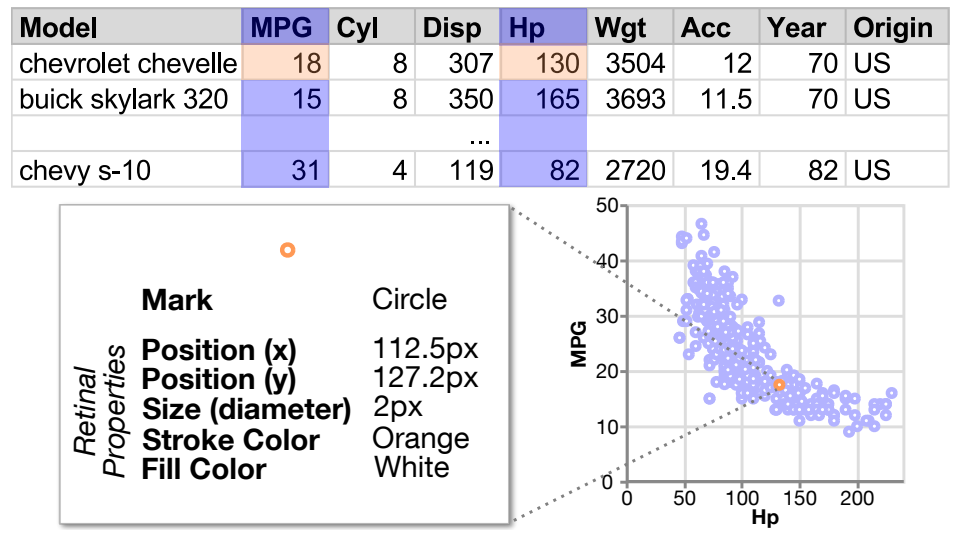}
   \caption{\label{fig:vizml}
     Example of underlying data and its visualization presentation~\cite{viznet}.}
     \vspace{-5mm}
\end{figure}

The foundation for designing and creating visualizations is the underlying data, as visualization serves as the medium for data presentation.
\removed{Structured data, such as tabular, network, and volume data, are particularly appealing to users.}
As a result, the underlying data is a crucial aspect of visualization datasets.
\revision{The most prevalent data formats are structured data, such as tabular, network, and volume data.}
For instance, tabular data is the most widely used format \removed{by users }and enables the application of deep learning techniques in visualization construction.
VizNet~\cite{viznet} is a large-scale corpus of 31 million underlying data obtained from open data repositories and online visualization galleries; the underlying data have an average of 17 records across 3 attributes.
Of the entire corpus, 51\% of dimensions recorded categorical data, 44\% recorded quantitative data, and the remaining dimensions recorded temporal data. VizNet serves as a common reference for training automated visualization tools and evaluating the effectiveness of visualization design, including comparing visualization techniques and developing benchmark models and algorithms for automated visual analysis.
\autoref{fig:vizml} illustrates a visualization and its corresponding underlying data.
Text-to-Viz~\cite{text2viz} provides a dataset of proportion-related statements and corresponding infographics consisting of 200 distinct infographic sheets and 983 infographic units. 
It categorizes the infographic units into several groups, including statistical-based, proportion-based, quantity-based, change-based, rank-based, timeline-based, process-based, and location-based infographics. The information from the infographics is extracted using CRF-based approaches. SNAP~\cite{leskovec2016snap} focuses on network and graph datasets, with a collection of over 50 large network datasets, ranging from tens of thousands to tens of millions of nodes and edges, including social networks, web graphs, road networks, internet networks, citation networks, collaboration networks, and communication networks. ChartSense~\cite{jung2017chartsense} provides the underlying data for chart images through semi-automatic, interactive data extraction algorithms that are optimized for each chart type.
In scientific visualization, volume data~\cite{avsarkisov_turbulent_2014, bartashevich_vector_2018} and flow data~\cite{kim_transport-based_2019, moitinho_gaia_2017} have also garnered interest among researchers.

\subsection{Visualization Components}
\begin{figure}[htb]
  \centering
  \includegraphics[width=\columnwidth]{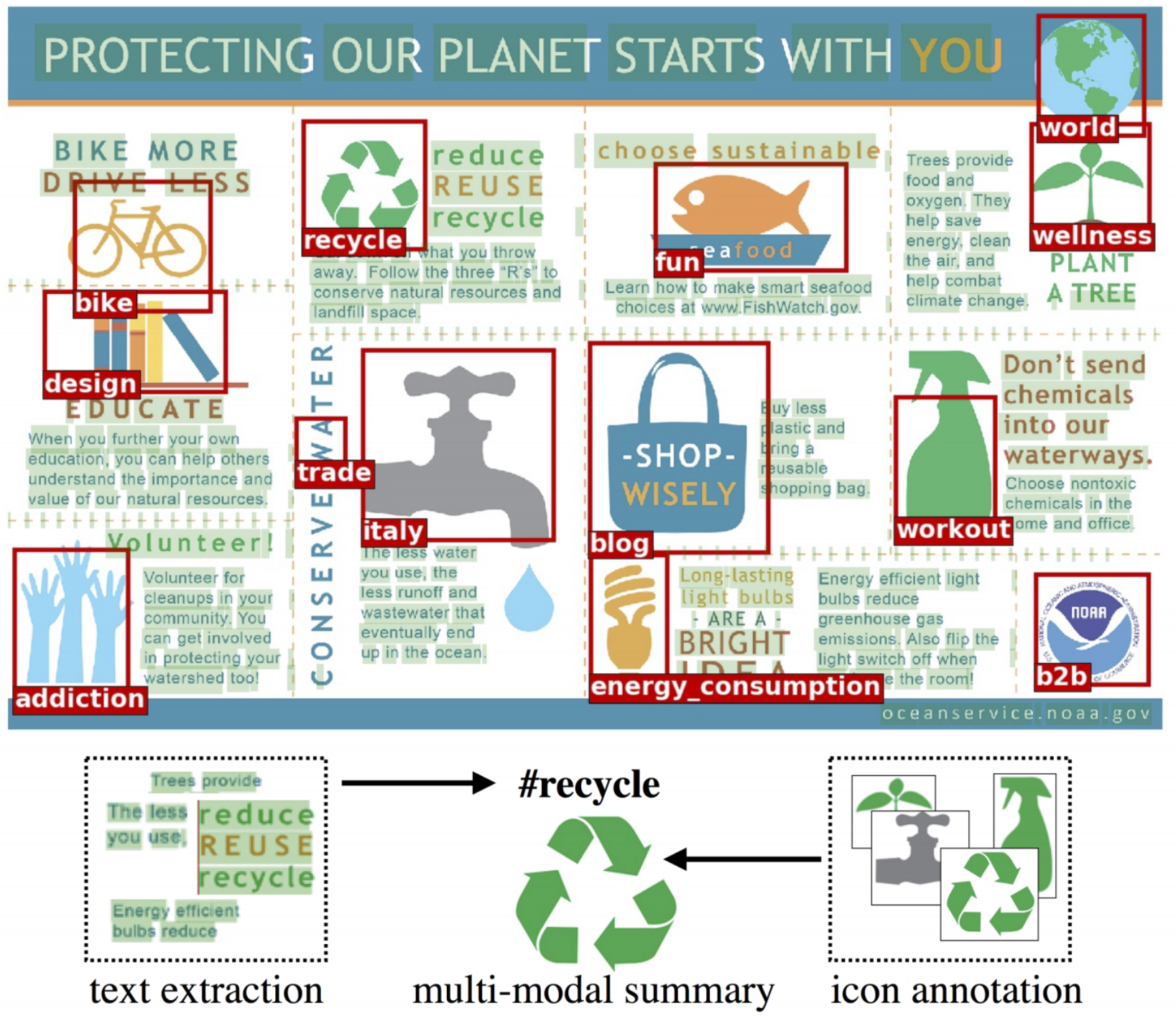}
   \caption{\label{fig:visually29K}
     The output of the fully-automatic annotation system in visually29K~\cite{Madan2021}.}
     \vspace{-5mm}
\end{figure}

Visualization components are constituents of a visualization presentation with certain semantics and functions, including visual marks and other auxiliary visual elements.
\revision{For image-based visualizations, datasets are obtained from visualization presentations through manual annotation or by using computer vision techniques such as image segmentation and semantic segmentation.
For specifications-based visualization, the content can be directly parsed according to the rules.}
There have been some approaches~\cite{reverse, lai2020automatic} that labeled visualization components (e.g., legends and axes) with trained deep-learning models. REV~\cite{reverse} merged the images from three chart corpora: Vega Charts (VEG), Quartz news website (QTZ)~\footnote{Quartz news website: \url{https://qz.com/}}, and Academic Paper Figures (ACA) for a total of 5,125 images.
The reverse-engineering visualizations recovered visual encoding from chart images using inferred text elements. This work detects textual elements in charts, classifies their role (e.g., chart title, X-axis label, Y-axis title, etc.), and recovers the textual content using the optical character recognition (OCR) technique. With the identified text elements and graphical mark type, the encoding specifications of an input chart image are automatically inferred. Visually29K~\cite{Madan2021} collects infographics with stylistically and semantically diverse visuals from the Visually design website and textual elements and labels stand-alone visual elements in infographics using a trained object detector, including the icon positions and semantics. As shown in~\autoref{fig:visually29K}, each infographic is annotated with 1-9 tags out of a set of 391 tag categories. Automatic-Annotation~\cite{reverse, lai2020automatic} identified and extracted visual elements in the target visualization, along with their visual properties, with a Mask R-CNN model. This work also extracts texts in the chart using optical character recognition (OCR) and parses the textual description using Natural Language Processing (NLP) engines. Chen et al.~\cite{timeline} provided a synthesized timeline dataset and a collected real-world timeline dataset from the Internet. The information in the bitmap timelines is deconstructed using a multi-task deep neural network, including the global information, i.e., the representation, scale, layout, and orientation, and the local information, i.e., the location, category, and pixels of each visual element on the timeline. Lu et al.~\cite{infovif} annotated the positions and types of important elements in infographics to provide information on their structure. 

\begin{figure}[htbp]
  \centering
  \includegraphics[width=\columnwidth]{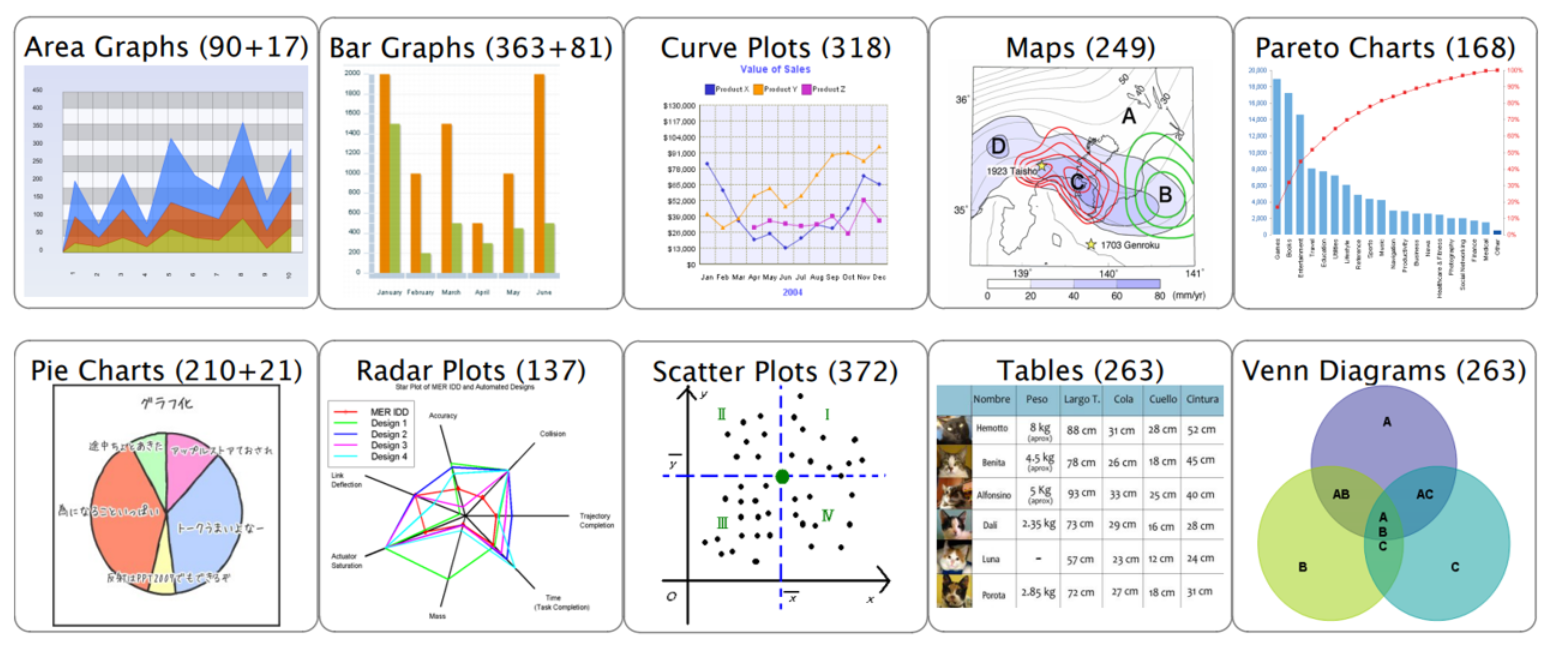}
   \caption{\label{fig:revision}
     The 10-category chart image corpus of Revision~\cite{revision}.}
\end{figure}

\subsection{Visualization Presentation}
A visualization presentation can come in either raster or vector format and can be either static or animated, with the option of being a single-view or multi-view representation.
These visualizations have various purposes, including exploring visualization techniques, conducting user experiments on visual perception, and training machine learning models with the underlying data.
Approaches, such as ReVision~\cite{revision}, ChartSense~\cite{jung2017chartsense}, and MASSVIS~\cite{memorable}, have collected static, single-view raster visualizations.
ReVision~\cite{revision} employs computer vision and machine learning to identify chart types, such as pie charts, bar charts, and scatterplots, in a bitmap image and creates an interactive gallery of redesigned charts using principles of perception-based design (as shown in Figure~\ref{fig:revision}). ChartSense~\cite{jung2017chartsense} uses a deep learning-based classifier to determine the chart type, which can then be further analyzed or improved for better perception. MASSVIS~\cite{memorable} collects a dataset of 2,070 single-panel visualizations from sources like news media, government reports, scientific journals, and infographics, categorizing them by type (e.g., bar charts and line graphs) and annotating them with attributes, such as data-ink scale and visual density ratings. The visualizations are also assessed for memory scores through Amazon's Mechanical Turk to identify the key factors that influence the retention and forgetting of visualizations.
Beagle~\cite{beagle} collects visualizations in vector graphics format. In addition to static, single-view visualizations, multi-view~\cite{chen2020composition} and animated visualizations~\cite{datagif, shi2021autoclips} have also been collected and studied.

\subsection{Additional Information}

\begin{figure}[htb]
  \centering
  \includegraphics[width=\columnwidth]{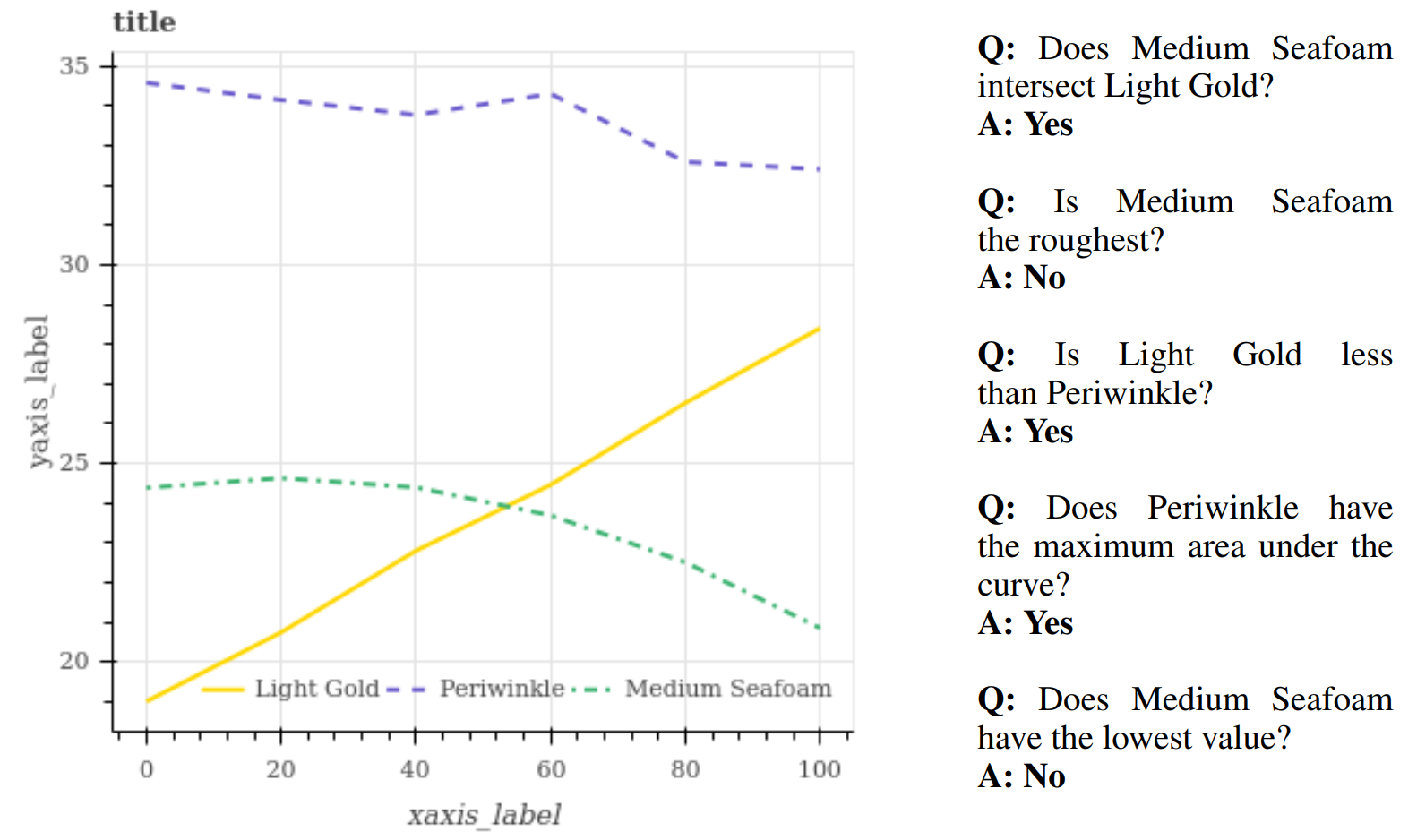}
   \caption{\label{fig:figureqa}
     Sample line plot figure with question-answer pairs in FigureQA~\cite{figureqa}.}
\end{figure}

The information that extends beyond the construction of a visualization includes user queries, feedback, and comments. This information may be incorporated into the visualization in the form of descriptions and annotations or may be separate from the visual presentation as the natural language interface and human perceptions.
FigureQA~\cite{figureqa} is a visual reasoning dataset that includes questions and answers related to charts, with over one million question-answer pairs based on over 100,000 images. The images are synthetic and depict scientific-style figures belonging to five classes, including line plots, dot-line plots, vertical and horizontal bar graphs, and pie charts. \autoref{fig:figureqa} shows an example of FigureQA.
The dataset features multiple plot elements and information synthesized spatially throughout the figures, for instance, bounding-box annotations for all plot elements, which can be used to train models to recognize patterns in visual representations of data. Some datasets also capture users' perceptions and understanding of visualizations.
Srinivasan et al.~\cite{srinivasan2021collecting} collected a corpus that maps natural language utterances to visualizations, with utterances labeled by over one hundred participants in a series of ten visualizations for a given dataset. This corpus can be used to evaluate existing natural language interfaces (NLI) for data visualization and to develop new systems and models to generate visualizations from natural language utterances. Moreover, human perception is a crucial aspect of visualization, such as users' recognition~\cite{memorable} and ratings of aesthetics~\cite{harrison2015infographic}.

\section{Why: Usage of Dataset}
\begin{figure}[htb]
  \centering
  \includegraphics[width=\columnwidth]{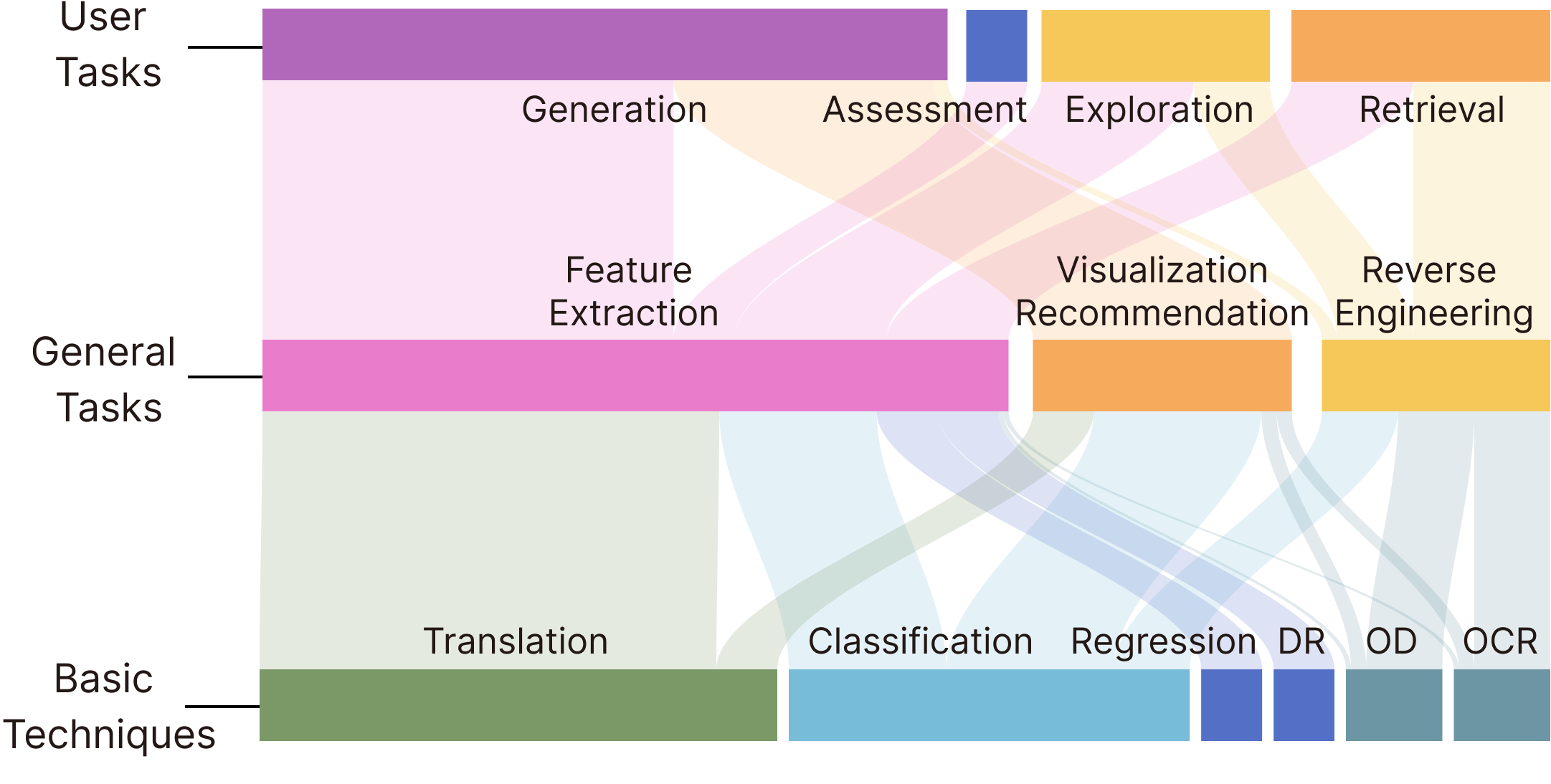}
   \caption{\label{fig:why}
    We organized the usage of visualization datasets into a three-layer structure, including basic machine learning techniques, general tasks, and user tasks (from bottom to top).
	Each row represents a layer, and each rectangular node represents papers belonging to that type. Additionally, each strip between nodes represents papers belonging to both types.
	\revision{The width of the node and strip is proportional to the number of papers.}
	}
\end{figure}

The integration of machine learning into visualization has enhanced its generation, retrieval, exploration, and assessment.
To implement machine learning, visualization datasets are required as training data, \revision{which vary in dataset components and data pairs for different tasks.}
We analyzed the tasks and methods presented in current works, and organized the utilization of datasets into a three-layer structure, as illustrated in~\autoref{fig:why}, encompassing basic techniques, general tasks, and applications. This section will examine the features of each level and the connections between the different levels.

\subsection{Basic Techniques}

In the field of machine learning, there are many common basic techniques such as classification, regression, object detection, machine translation, OCR (Optical Character Recognition), and dimension reduction.
For each technique, various machine learning algorithms have been proposed and trained on different datasets to obtain a model that meets the requirements.
In visualization, these machine learning algorithms can be used as a means of analyzing and understanding visualization and the results obtained can be used for subsequent visualization recommendation, reverse engineering, and feature extraction tasks.

\textbf{Classification} identifies which group or category a sample belongs to based on some features or characteristics of the data.
In the field of machine learning, algorithms such as logistic regression, support vector machine, random forest, decision tree, deep learning, etc. can be used to solve classification tasks.
In visualization, classification tasks include classifying visualization chart types, such as ReVision~\cite{revision}, Vizbywiki~\cite{lin2018vizbywiki}, ChartSense~\cite{jung2017chartsense}, and FigureSeer~\cite{siegel2016figureseer}. It also includes classifying symbols and text in visualizations, with FigureSeer~\cite{siegel2016figureseer} and ReVision~\cite{revision} using convolutional neural networks to classify mark types, and FigureSeer~\cite{siegel2016figureseer} training a classifier to determine the text role as a legend label. Additionally, it includes classifying the data in visualizations, with Text-to-Viz~\cite{text2viz} training a text analyzer to extract segments from natural language statements and generate infographics. Sherlock uses a deep neural network to classify semantic types on data columns from the VizNet~\cite{viznet} corpus, which can better aid in data mining and visualization.

\textbf{Regression} is a supervised task that predicts the value of a target variable based on the input of independent variables.
Traditional regression methods include linear regression, ridge regression, random forests, and deep learning methods.
In visualization, regression tasks include evaluating and scoring visualization.
\revision{For example, the memorability and aesthetics of visualized images are evaluated~\cite{fu2019visualization}, and the analysis and evaluation of graph visualization are performed~\cite{Haleem2018EvaluatingTR, cai2021AMachine}.}
In recent years, the evaluation of graph visualization has been a topic of interest, and machine learning techniques have been employed to tackle the problem. Various readability metrics have been proposed to model the layout of graph visualization. However, computing these metrics can be time-consuming, especially for large and dense graphs. To address this issue, Haleem et al.~\cite{Haleem2018EvaluatingTR} proposed a deep learning-based approach for directly evaluating the readability of graph layouts from images.
They designed a convolutional neural network architecture, trained it on a benchmark dataset of graph layout images, and generated readability metrics like node occlusion, edge crossing, and group overlapping. There is a connection between the readability metrics and the user's preference for a graph layout. Cai et al.~\cite{cai2021AMachine} leveraged the readability metrics to train a deep learning model using transfer learning methods, which can predict people's preferences for a graph layout.
For scientific visualization, there are two categories: predicting the final rendering results and predicting the rendering parameters needed to improve the results.
Rendering parameters and corresponding results are widely used as data pairs for scientific visualization datasets for learning methods. Those two are used as input and predicted targets for various tasks, including rendering synthesis~\cite{he2019insitunet, hong2019dnn}, occlusion prediction~\cite{engel2020deep}, super-resolution~\cite{han2021stnet, han2022coordnet}.
Completing regression tasks can effectively reduce rendering costs and time consumption and improve rendering quality by using learning methods appropriately.

\textbf{Dimensionality reduction} algorithms transform data from high-dimensional spaces to low-dimensional spaces, preserving meaningful attributes of the original data. Algorithms such as principal component analysis, t-SNE, and VAE~\cite{cinelli2021variational} are commonly used for dimensionality reduction. In visualization, representing visualization images as high-dimensional vectors facilitates the reconstruction of visualizations and exploration of the design space. Visualization Assessment~\cite{fu2019visualization} uses VAE for dimensionality reduction of visualization images for the reconstruction of new visualizations. ChartSeer~\cite{Zhao2020a} and GoTreeScape~\cite{gotreescape} use variational autoencoders to generate high-dimensional vector representations of graph visualization.

\textbf{Translation} is a process of using machine learning algorithms to translate text from one language to another.
In visualization, the translation task can be abstracted as a process of transforming one object into another.
For example, Dibia et al.\cite{data2vis} view the visualization creation process as a direct generation of specification files for visualization from data.
Luo et al.\cite{luo2021synthesizing} translate the user-given natural language into structured visualization queries, and then generate the desired visualization that meets the requirements.

\textbf{Object Detection} algorithms are a category of computer vision algorithms that aim to identify objects in images or videos and annotate them with bounding boxes. Deep learning-based target detection algorithms such as Fast R-CNN~\cite{fastrcnn} and YOLO~\cite{Redmon2018YOLOv3AI} have shown remarkable performance, providing great assistance for computer understanding of images.
These algorithms are extremely important in the field of computer vision, as they can be applied to a wide range of applications such as autonomous driving.
\revision{In visualization, the use of target detection algorithms can detect and extract the positions of visual elements and function components (e.g., legends, axes, and titles) in visualization images.}

\textbf{OCR} (Optical Character Recognition) is a technology that enables the recognition of text from images or scanned documents and its conversion into editable electronic text.
There are OCR algorithms based on rules, template matching, and deep learning.
In visualizations, to obtain the original information contained in the visual elements, approaches~\cite{timeline, lai2020automatic} often extract text content such as titles, captions, and texts in the axes.
Once the text is extracted and processed, the information can be used to further reconstruct the visualizations.

\subsection{General Tasks}
Building upon basic machine learning techniques, general tasks of visualization can be constructed. The pipeline of visualization highlights the importance of visualization recommendations, including the selection of appropriate data attributes and visual mapping. Conversely, reverse engineering aims to restore this information from existing visualizations. Furthermore, another general task is feature extraction, which extracts features from objects such as data and visualizations.

\textbf{Visualization recommendation} refers to the recommendation of suitable visualizations based on data and user intent, including the selection of data attributes and visual mappings. According to the basic techniques used in the recommendation, existing works can be divided into three categories. The first is the classification-based recommendation. For example,  Hu et al.~\cite{vizml} predicted the chart type of visualizations based on the features of data tables. The second is the regression-based recommendation, such as DeepEye~\cite{deepeye}, which enumerated the combination of data attributes and visual mappings, scored them using regression methods, and recommended the best. The third is the translation-based recommendation, such as translating data directly into a visualization specification~\cite{data2vis} or translating natural language into structured visualization queries~\cite{luo2021synthesizing}.

\textbf{Reverse engineering} refers to the process of extracting data and encoding information from visualizations. A common pipeline consists of three steps~\cite{choi2019visualizing, reverse}: deconstructing visualization images, recovering the visual mappings, and extracting underlying data. The deconstruction of visualization components involves detecting and categorizing various components in the visualization image, such as graphics, axes, and legends.
After deconstruction, the type of chart needs to be predicted through classification, and visual mapping needs to be inferred from components such as axes and legends through heuristic rules. Finally, based on the visual mapping, the underlying data can be extracted by regression tasks or position information of graphics. 

\begin{figure}[htb]

	\centering
	\subfigure[]{
	   \centering
		\begin{minipage}[b]{0.99\columnwidth}
			\includegraphics[width=0.99\columnwidth]{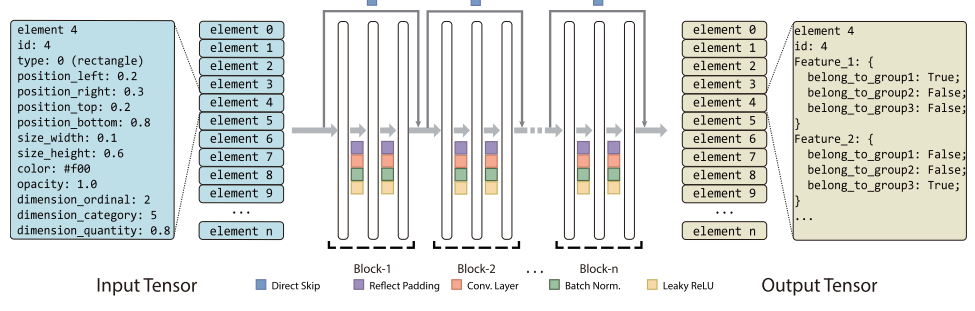}
		\end{minipage}
		\label{fig:feature_extraction_explicit}
	}
    \subfigure[]{
	   \centering
    	\begin{minipage}[b]{0.99\columnwidth}
			\includegraphics[width=0.99\columnwidth]{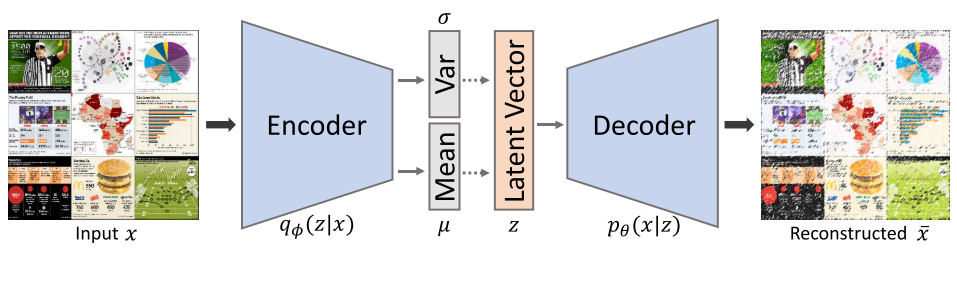}
    	\end{minipage}
		\label{fig:feature_extraction_implicit}
    }
	\caption{Feature extraction: (a) Extracting data features explicitly in AutoCaption~\cite{liu2020autocaption}, which generates descriptive feature expressions from visual elements; (b) Extracting hidden space vectors of infographics implicitly, through reconstructing the images by Autoencoders~\cite{fu2019visualization}.}
	\label{fig:feature_extraction}
\end{figure}

\textbf{Feature extraction} refers to the process of extracting features from objects (e.g., visualizations, data, visual components).
Based on the form of the feature, it can be divided into explicit feature extraction and implicit feature extraction.
Explicit feature extraction represents the features explicitly as categories, numerical values, and texts, corresponding to the basic supervised techniques such as classification, regression, and translation.
For example, Liu et al.~\cite{liu2020autocaption} built a dataset that contains data, visual elements, and descriptive feature expressions to extract data features (\autoref{fig:feature_extraction} (a)).
\removed{Implicit feature extraction represents the features as hidden space vectors and is an intermediate result in basic machine learning techniques.}
\revision{Implicit feature extraction entails the autonomous identification of suitable data representations by a deep model. For example, an autoencoder architecture functions by encoding data into a condensed vector form and subsequently reconstructing it. Through rigorous training, the vectors within this latent space are refined to effectively encapsulate the essential characteristics of the original data form.}
For instance, Fu et al.~\cite{fu2019visualization} used a dimensionality reduction method based on VAE to learn vectors from a large number of infographics images (\autoref{fig:feature_extraction} (b)).
When data labels are available, supervised machine learning can be used to generate implicit space vectors. 
For example, Lee et al.~\cite{lee2021viral} used a supervised trained chart classification model to extract vectors from COVID-19 visualizations.
In the absence of data labels, dimensionality reduction methods can be used.

\subsection{\revision{User Tasks}}

\removed{Visualization datasets exhibit a varied range of applications in diverse scenarios.
Typically, the problem we face is to gain insights from data through visualization, which the generation of visualizations, captions, and answers or retrieval from pre-existing visualizations can accomplish. 
In some other cases, analyzing and exploring new forms and utilization of visualization are also imperative.
For this purpose, exploration and assessment of existing designs are needed.}
\revision{An information visualization process comprises data transformation, visualization encoding, and visualization rendering. Once visualization results are obtained, authors create descriptions, titles, or implement question-answering methods to aid user understanding. Deep learning approaches can facilitate parts of the process from data to visualization via supervised fitting.
Furthermore, with a set of visualizations, one can retrieve specific visualizations, explore the set, or conduct evaluations based on it.}

\textbf{Generation} refers to the creation of new information carriers from existing ones.
\revision{In the visualization area, generation encompasses the generation of charts, volume rendering results, responses to questions, and the production of annotations and captions.}
The generation of charts involves recommending appropriate visual mappings based on a combination of data attributes.
This is typically performed through automatic generation from data, either by iterating through different combinations of data attributes and visual mappings and evaluating their effectiveness~\cite{deepeye}, or by directly recommending a visualization specification based on the data~\cite{data2vis, vizml, Wu2022}.
A large number of charts can be generated from a given dataset through these methods, but the user often needs to manually inspect them to determine which ones best match their intention.
To integrate the user's intention more effectively into the generation process, some research has proposed generating charts from natural language, which involves extracting tasks, data attributes, and values from natural language statements and recommending visual representations accordingly. For example, Text-to-Viz~\cite{text2viz} extracts data entities related to proportion from natural language statements and recommends infographics, while Luo et al.~\cite{luo2021synthesizing} convert natural language into structured visualization queries and recommend charts.
Generation of volume rendering results refers to the process of synthesizing volume rendering results through parameters such as viewing points or enhancing the original results. Starting from the parameters, the volume rendering result can be directly generated. The input parameters can be the viewing point and the transfer function~\cite{berger2018generative}, as well as the simulation parameters and visualization parameters~\cite{he2019insitunet}. Enhancements to the original volume rendering results include super-resolution~\cite{han2021stnet,han2022coordnet} and predicting ambient occlusion~\cite{engel2020deep}.
\revision{Han et al.~\cite{han2021stnet} create high-resolution images from low-resolution volume rendering images. Engel et al.~\cite{engel2020deep} use a deep learning model to predict ambient occlusion based on the opacity volume derived from the intensity volume and the transfer function.}

\removed{
However, when users ask analytical questions, it is difficult to provide answers through simple natural language, such as inquiring about the correlation between two attributes.
To address this, some research has proposed generating visualization results based on data and natural language.
Another type of research generates natural language answers based on visualization and natural language.
However, these answers only provide the answer without explaining the reasoning process, raising doubts about their credibility among users.
To address this issue, Kim et al. proposed to describe the reasoning process through natural language in addition to the answer.}
\revision{
\textbf{Question-answering} is the process of obtaining information from data and visualizations by asking questions.
In the field of visualization, some research generates natural language answers based on visualization and natural language~\cite{figureqa, Kafle2018, chartqa}. However, when users pose analytical questions, it can sometimes be challenging to provide answers through simple natural language, such as when inquiring about the correlation between two attributes. To tackle this, certain approaches generate visualization results based on underlying data and natural language questions~\cite{liu2021advisor, luo2021synthesizing}.
Unlike generating visualizations from underlying data, some methods~\cite{Kim2020, lai2020automatic} focus on question-answering (QA) based on existing visualizations. For example, Kim et al.\cite{Kim2020} suggested describing the reasoning process in natural language alongside the answer. Lai et al.\cite{lai2020automatic} determine the content to be annotated using natural language, and extract text and graphic entities for annotation through feature extraction and reverse engineering before creating annotations by linking them, which also represents a form of QA based on existing charts.}

\begin{figure}[htb]

	\centering
    \begin{minipage}[b]{\columnwidth}
        \includegraphics[width=\columnwidth]{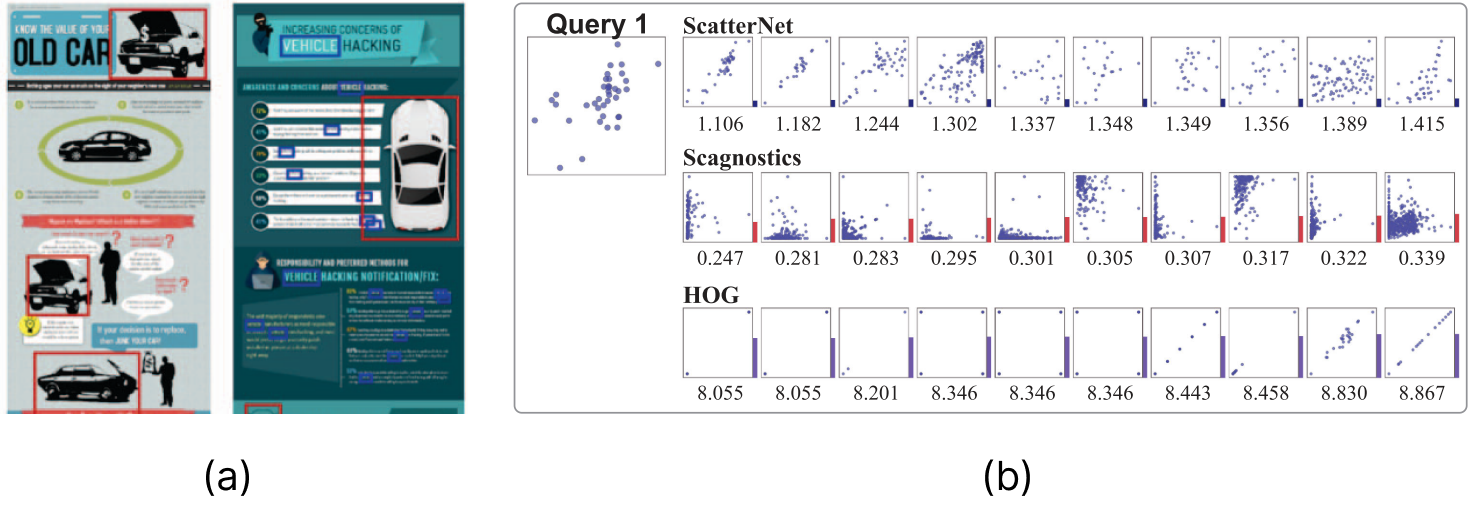}
    \end{minipage}
	\caption{Retrieval: (a) searching by matching, which retrieves infographics by hashtags\revision{\cite{Madan2021}}; (b) searching by the similarity of hidden space vectors in ScatterNet~\cite{ma2018scatternet}, which is used to retrieve similar scatterplots.}
	\label{fig:retrieval}
\end{figure}

\textbf{Retrieval} refers to the process of searching for a subset that meets certain criteria from a collection of information carriers. This can involve searching for visualizations that match certain conditions in a collection of visualizations. The input for the query can include visualization images~\cite{ma2018scatternet}, data attributes and visual mappings~\cite{lin2018vizbywiki,chen2015diagramflyer,siegel2016figureseer}, or themes~\cite{vizcommender,Madan2021}. Once the query input is specified, a direct approach is to match it~\cite{chen2015diagramflyer, Madan2021}. For example, Madan et al.~\cite{Madan2021} parse text and icon information from infographics through reverse engineering and support keyword-based retrieval (\autoref{fig:retrieval}(a)). The disadvantage of this matching approach is that the users need to specify the query input accurately. Otherwise, they may miss the desired results. To address this issue, some works propose representing the existing query input as hidden space vectors and calculating similarities, such as ScatterNet~\cite{ma2018scatternet}, which learns hidden space vectors based on the similarity of scatterplots and uses them for retrieval (\autoref{fig:retrieval}(b)).

\begin{figure}[htb]
	\centering
    \begin{minipage}[b]{0.4\textwidth}
        \includegraphics[width=0.99\columnwidth]{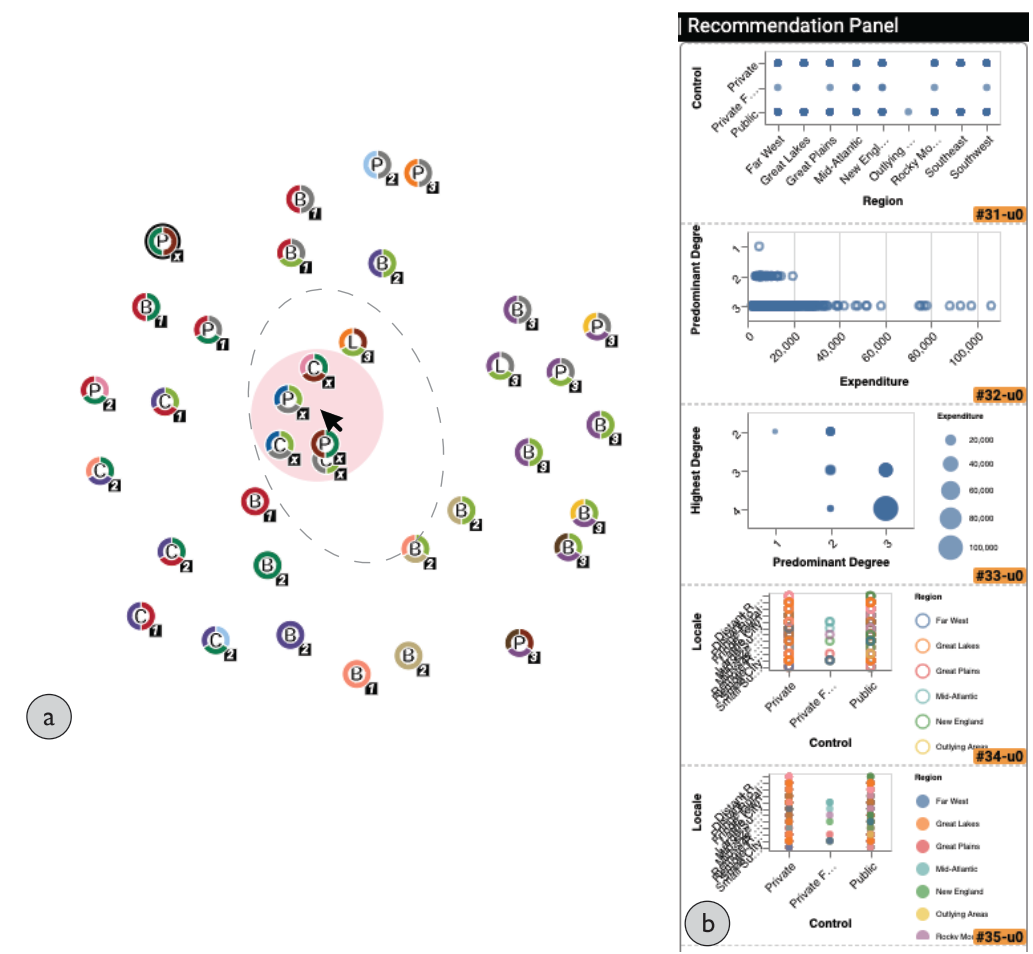}
    \end{minipage}
	\caption{\label{fig:exploration}Exploration: Projection analysis with recommendations using the hidden space vectors of charts in ChartSeer~\cite{Zhao2020a}.}
\end{figure}

\textbf{Exploration} refers to the examination of data sets to uncover valuable insights. One category is the analysis of visualization usage, such as Beagle~\cite{beagle}, which crawled a significant number of charts from the web, obtained the chart types through explicit feature extraction, and analyzed the prevalent chart types found on the internet. 
\removed{On the other hand, }Madan et al.~\cite{Madan2021} parsed categories, positions, and related text information of icons from infographics, and synthesized themes and icons.
Another category is the exploration of the design space of visualizations, where users need to invest a considerable amount of time in examining the vast space due to the abundance of options for data attributes, data transformation methods, and design dimensions. To support users in discovering relevant visualizations from the pool, some works extract the hidden space vectors of charts through feature extraction methods, which are then utilized for subsequent projection analysis and recommendations. For example, ChartSeer~\cite{Zhao2020a} constructed a data set of a large number of Vega-Lite~\cite{satyanarayan2016vegalite} configuration files by traversing data attributes and design dimensions to learn the hidden space vectors and assist users in identifying suitable charts (\autoref{fig:exploration}). In addition to basic charts, there are explorations for tree visualizations~\cite{gotreescape} and volume rendering~\cite{shen2022idlat}.

\begin{figure}[htb]

	\centering
    \begin{minipage}[b]{0.99\columnwidth}
        \includegraphics[width=0.99\columnwidth]{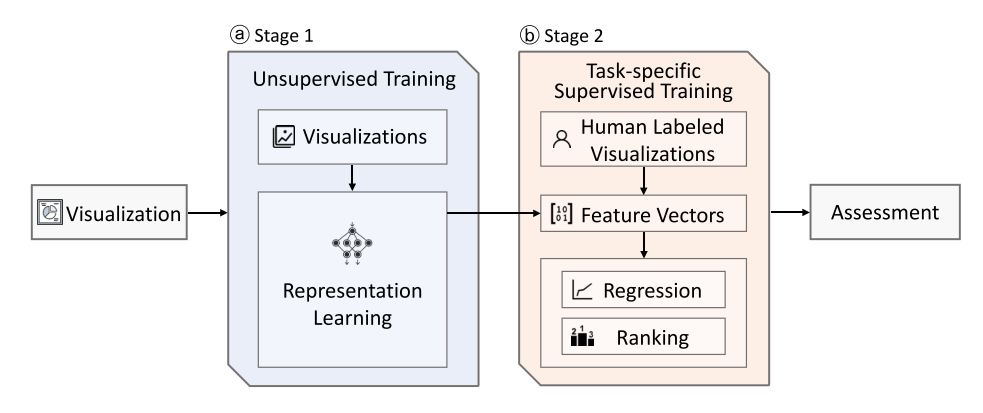}
    \end{minipage}
	\caption{\label{fig:assessment}Assessment: Training a model using both a large-scale infographic dataset and a small dataset of infographics with aesthetics labels~\cite{fu2019visualization}.}
\end{figure}

\removed{\textbf{Assessment} refers to the evaluation of the performance of visualization based on specific metrics.} 
\revision{\textbf{Evaluation} of visualization have been conducted from various efforts, including the effectiveness of visual encoding~\cite{viznet}, readability~\cite{Haleem2018EvaluatingTR}, user preference~\cite{cai2021AMachine}, and aesthetics~\cite{fu2019visualization}.}
A straightforward approach is to use the metrics values as inputs and outputs for models trained through explicit feature extraction. For instance, Hu et al.~\cite{viznet} leveraged the visualization effectiveness and encoding effectiveness to learning effectiveness. However, this approach may not perform well when the dataset is small. To address this issue, some works propose to obtain latent space vector representations of visualizations through feature extraction, which can be used as inputs for the above methods. For example, Fu et al.~\cite{fu2019visualization} trained a model using both a large-scale infographic dataset and a small dataset of infographics with aesthetics labels (\autoref{fig:assessment}).
Another solution is to first train a model using data generated by algorithms and then fine-tune it using human-labeled data. Cai et al.~\cite{cai2021AMachine} used both readability metrics and user preference data to train a deep learning model for predicting people's preferences for graph layouts.

\subsection{Findings}
We analyzed the types of existing papers across different layers (\autoref{fig:why}), and figured out some key findings. 
The most commonly observed application is generation, followed by retrieval, exploration, and evaluation. For general tasks, feature extraction is the most prevalent, followed by visualization recommendation and reverse engineering. Among the fundamental machine learning methods, translation appears most frequently, followed by classification, object detection, OCR, regression, and dimensionality reduction.
In terms of the relationships between different layers, several significant patterns emerge. First, the process of generation is primarily accomplished through feature extraction. This involves utilizing translation methods to extract relevant features from the data and subsequently generating visualizations.
Second, when it comes to visualization recommendation and generation, classification techniques play a crucial role. These techniques are employed to classify and categorize visualization data, enabling the system to recommend suitable visualizations. 
Third, for exploration and retrieval purposes, feature extraction plays a key role, and it is predominantly achieved through classification techniques. 
Fourth, in the context of reverse engineering for exploration and retrieval, multiple techniques are used. This involves the use of classification techniques, along with object detection and OCR techniques.

\section{How: Construction of Dataset}

\begin{figure}[htb]
  \centering
  \includegraphics[width=\columnwidth]{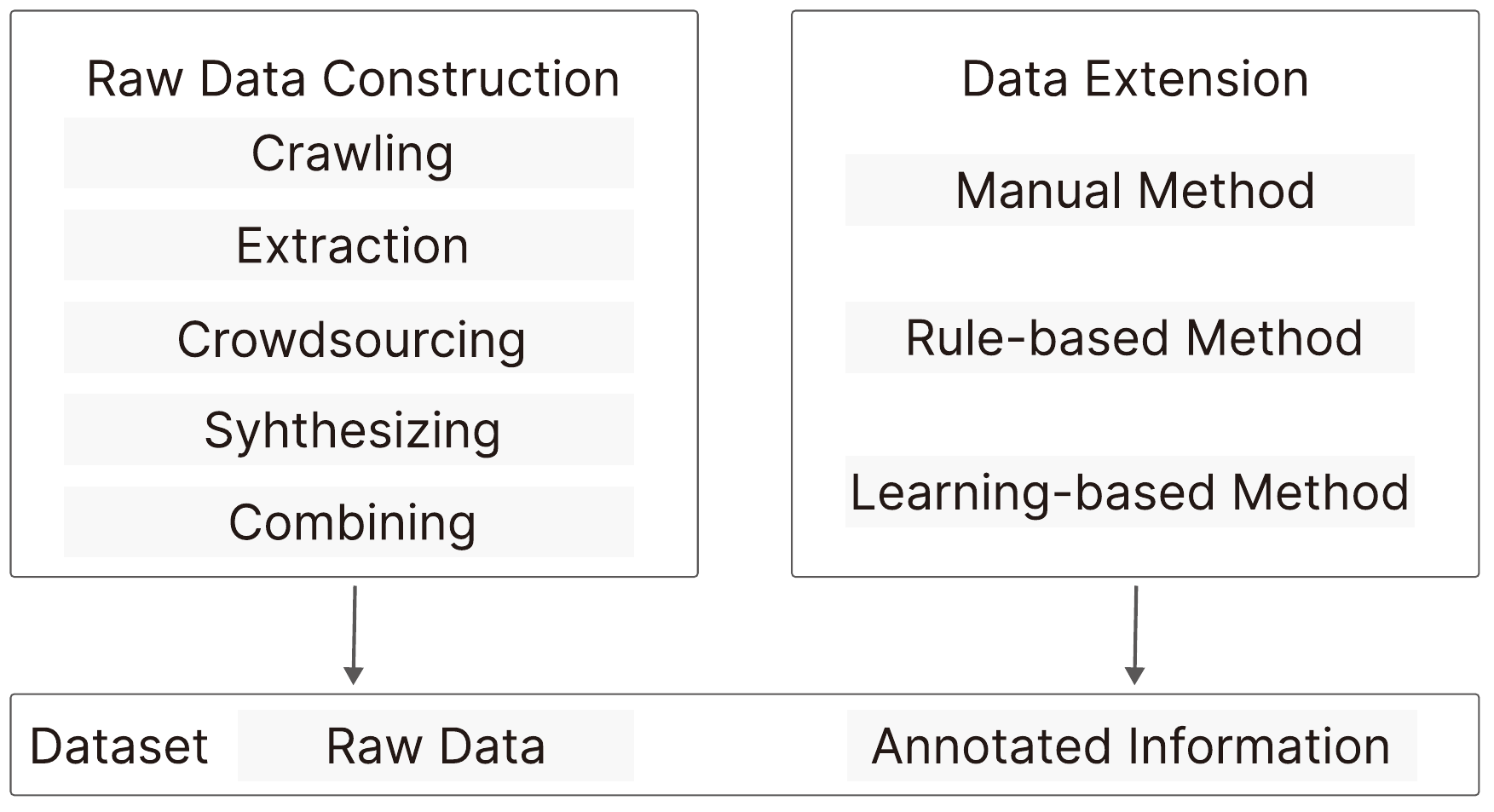}
   \caption{\label{fig:how}
     Dataset construction consists of raw data construction and data augmentation obtaining detailed information from visualizations.}

\end{figure}

Constructing visualization datasets includes raw data construction and data augmentation, as shown in ~\autoref{fig:how}. 
\revision{\textbf{Raw data} means data information that can be directly obtained from documents, usually with only one type of data information, including underlying data like data tables or graph data, and visualizations in the formats of raster images, and vector images.}
Processing of raw data calculating derived information to be included in the datasets.
The annotation of raw data can include detailed information related to visualizations and human perception.

\subsection{Raw Data Construction}

The raw data mainly comes from five sources, namely, crawling, extraction, crowdsourcing, synthesizing, and combining.
Some approaches~\cite{covid19landscape} collect the data by crowdsourcing, utilizing the power of the general public.
Synthesizing~\cite{figureqa, Kafle2018} dataset can construct a large scale of the dataset.
The limitation of the synthesized dataset is that it lacks diversity in underlying data, visual encoding, and style. 
Therefore, other works directly collect desired data types (e.g. visualizations) from the real world like online websites, for example, Beagle~\cite{beagle} and ReVision~\cite{revision}.
Some visualizations are stored in documents, e.g., PDF files. Extraction methods are applied to collect visualizations in these media, to create visualization datasets, for example, VisImages~\cite{visimages}.

\textbf{Automatic web crawling} can be conducted either with or without a specified list of websites.
The list may include multiple centralized websites, such as news media websites that provide data news, visualization blogs, government agencies, or international organizations that publish data reports.
For instance, MASSVIS~\cite{memorable} crawls visualizations from various websites across different fields automatically. D3 search~\cite{d3search} crawls D3-based visualization collection sites, while VizML~\cite{vizml} crawls data and visual encoding specifications from the Plotly website~\cite{plotly}.
InfoVIF~\cite{infovif} collects infographics from two infographic design websites.
Another centralized collection approach utilizes search engine sites, such as video search on video sites like YouTube and Vimeo, and image search functions on sites like Google, Flickr, Twitter, and Reddit. This approach is widely used in the field of natural images; for example, ImageNet~\cite{imagenet} searches for images and tags using keywords from WordNet~\cite{wordnet}. In the visualization domain, the dataset for training automatic annotation models~\cite{lai2020automatic} is collected by using visualization types as keywords for crawling with search engines. Shi et al.~\cite{shi2021video} also use video search engine results as one of their data sources when collecting data videos.
The advantage of web crawling is that it can involve visualizations used and disseminated in reality, and novel visualizations can be collected to construct the visualization datasets. However, it also has disadvantages, such as the difficulty in identifying websites, the similarity of visualizations from the same website, and the potential for bias in the analysis results due to the way of specifying data sources. Additionally, appropriate crawling and web structure analysis strategies need to be developed for different websites, and the selection of keywords and filtering of visualization images must be considered for search engine types. This method is limited in the amount of visualization data collected and may also suffer from bias due to the limitations of search engines and recommendation algorithms. The automatic approach requires a special design to filter relevant images for visualization, and although some existing work can be automated, the dataset collection process is often performed with manual review.

\textbf{Automatic extraction from static documents} is a valuable technique for acquiring \revision{visualization-related components}, including visualization diagrams and descriptions.
One example of static documents are PDF files.
Parsing PDF files enables the analysis of academic papers, utilizing sophisticated toolkits capable of accurately determining the position of text and images.
Several research initiatives, including DiagramFlyer~\cite{chen2015diagramflyer}, FigureSeer~\cite{siegel2016figureseer}, SCICAP~\cite{hsu-etal-2021-scicap-generating}, and VizioMetrix~\cite{lee2016viziometrix}, have employed automatic extraction and visualization methods to obtain visualization and extract underlying data. Additionally, multi-view visualization datasets, such as those collected by Chen et al~\cite{chen2020composition} and VisImages~\cite{visimages}, are extracted from professional visualization papers and are considered to be of high quality due to their design and evaluation by experts.
Other files like Excel spreadsheets may also contain data and visualizations and can be used to construct visualization datasets~\cite{luo2021chartocr}.
\revision{However, the challenge of extracting visualization datasets from static documents includes the diverse formats and increased error in the extraction process.}
\removed{as compared to data collected from web pages.}

\textbf{Crowd-Sourcing collection} has the potential to improve the quality of data collection from decentralized sources. 
In the case of the CrisisVis dataset~\cite{covid19landscape}, some of the visualizations were obtained through word-of-mouth and manually extracted and labeled by the authors. 
\removed{An alternative approach is to use automatic crawling methods, similar to how search engines operate, to freely crawl and extract Internet content.
For example, the authors of Beagle attempted the automatic crawling of charts but encountered issues with overwhelming duplicates, particularly in the case of automatically generated user portrait visualizations from StackOverflow. Although the distribution of visualizations on the Internet is uneven, making the method inefficient, one possible solution is to utilize existing Internet crawling projects, such as CommonCrawl, to extract visualizations. However, this approach still faces challenges in terms of efficiency and overhead, as well as difficulties in handling dynamically generated visualizations.}
SightLine~\cite{sightline} uses a browser plugin to extract visualizations and upload them to the server based on set heuristic rules as the user browses the visualization page. However, this approach faces challenges in finding users and may raise privacy concerns.
Sites such as Plotly~\cite{plotly} allow users to upload raw data and set up encoding mappings or select templates to generate visualizations.
Therefore, these sites collect a large amount of real raw data and user-defined mappings. The disadvantage is that these methods require a lot of time and other expenses and are suitable as a long-term collection tool, while it is difficult to guarantee the number of collections in the short term.
Moreover, these methods require a platform with a large influence to attract users.

\textbf{Synthesizing of Data and Visualization.}
Data and visualization can be synthesized based on rules to reduce the cost of collecting large volume datasets.
For example, FigureQA~\cite{figureqa} generates datasets for training QA tasks for visualization.
Synthesizing method is also often used in scientific visualization dataset construction.
Appropriate rendering techniques are required to obtain scientific visualizations. 
\removed{In this process, machine learning techniques can be applied to accomplish tasks related to image generation and processing. }
The data required by these learning models mainly includes paired information: rendering results (images) and corresponding parameters.
The most important part of constructing a scientific visualization dataset lies in the rendering parameters. When facing different tasks, the corresponding parameters needed by the learning models are also different.
However, in most cases, researchers will sample the corresponding parameter space based on rule-based methods to construct a visualization dataset.
For example, Berger et al.~\cite{berger2018generative} and Hong et al.~\cite{hong2019dnn} used the viewpoint and transfer function as parameters, and constructed datasets for parameter to rendering results pairs.
He et al.~\cite{he2019insitunet} sampled the ensemble data simulation parameter space, added parameters on the basis of viewpoint and transfer function, and constructed training datasets for ensemble scientific data.
Engel et al.\cite{engel2020deep} explored different opacity mappings and corresponding ambient occlusion in direct volume rendering results, and established a dataset for deep learning-based techniques for volumetric ambient occlusion.
In some cases, those rendering parameters are hidden behind datasets. Han et al.\cite{han2022coordnet} used a uniform coordinate representation to incorporate parameters for different learning tasks into the same framework, and designed a corresponding universal training network. In addition, Han et al.~\cite{han2021stnet} constructed a dataset for super-resolution, which only includes pairs of input low-resolution rendering results and targeted high-resolution results. However, they still traversed the rendering parameter space when constructing this training dataset. The corresponding generation parameters were not directly input to the adversarial network since super-resolution tasks do not require them.
Overall, the methods used in constructing training datasets in the field of scientific visualization are all based on sampling in the corresponding parameter space and obtaining visualization results using rendering methods based on those parameters. Thus, researchers can construct data pairs of parameters and render results as training data for deep learning models.
It is worth noting that work dedicated to using machine learning for tasks such as interpolation prediction and feature detection is not included as they focus on the underlying data instead of visualizations.

\textbf{Combination of Existing Datasets.}
Some work has been devoted to collecting data from the Internet Web, such as WebTables~\cite{cafarella2008webtables}, while there are also platforms that allow users to expose their data and visualizations like Plotly~\cite{plotly} and Tableau~\cite{tableaupublic}.
\removed{However, users upload data in a variety of formats that are difficult to use directly by analytic algorithms or visualization libraries.} \revision{Datasets from different sources can be combined together by converting into a unified format.}
VizNet~\cite{viznet} contains publicly available datasets from WebTables and Manyeyes and also provides cleaning of the data and calculation of important statistical values.

\subsection{Data \revision{Augmentation}}

Data augmentation consists of the process of augmenting existing components with additional information components from like visualization type, visualization component, etc.
Based on whether the augmentation is achieved by humans or machines, the augmentation method can be divided into three categories, including manual method (e.g.,, icon detection~\cite{Madan2021}, bounding box~\cite{visimages}, and natural language utterances~\cite{srinivasan2021collecting}), rule-based method, and learning-based method \removed{which leverages machine learning or deep-learning models to }get new information based on the knowledge distilled from training data.
Datasets can be extended by combining multiple augmentation methods to leverage their benefits.

\begin{figure}[htb]
  \centering
  \includegraphics[width=\columnwidth]{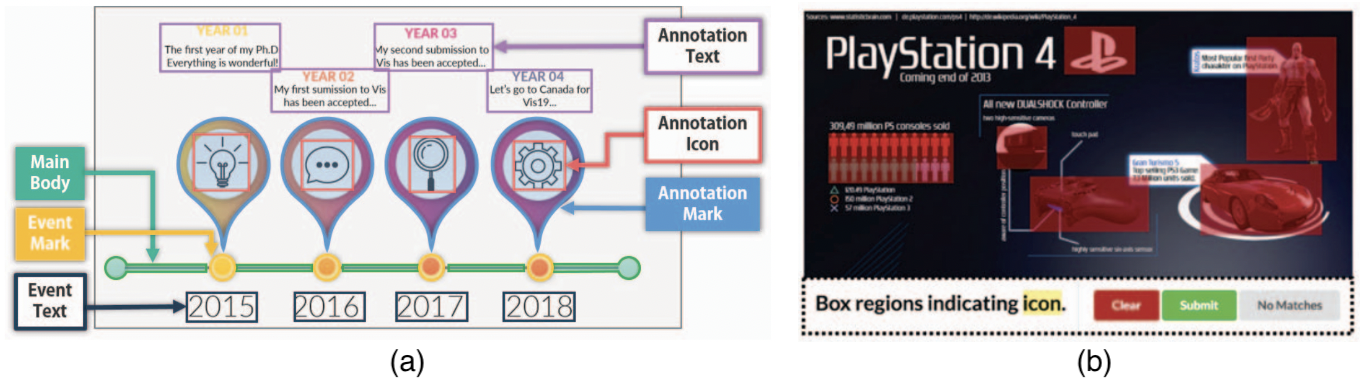}
   \caption{\label{fig:how_example}
     (a) Icon and text annotation in the Timeline dataset~\cite{timeline} (b) Datasets of icons in infographics proposed by Madan et al.~\cite{Madan2021}}
\end{figure}

\textbf{Manual Method.}
Datasets can be augmented by involving humans to label information including visualization types and component positions and types,
collecting perception data by asking participants to conduct tasks on visualizations, or asking people to describe visualizations with language.
Visualization-type information is obtained by human labeling in datasets such as Revision~\cite{revision}, MASSVIS~\cite{memorable}, and that of Choi et al.~\cite{choi2019visualizing}
Visual components in standard visualization charts are manually annotated with positions and component types~\cite{lai2020automatic, reverse}, and views in multi-view visualization systems are also annotated\removed{in}~\cite{chen2020composition}.
Infographics with more diverse styles are annotated by crowdsourcing to get information like icons~\cite{timeline, Madan2021}, texts~\cite{timeline}, and structural information~\cite{infovif}.
The result of icons and texts annotated in a timeline is illustrated in \autoref{fig:how_example} (1).
The part of the visualization dataset that pertains to perceptual cognition is typically obtained through user experiments.
Accuracy~\cite{draco, memorable} of completing tasks are commonly collected.
Human subject assessment of visualizations is also an important target to collect like aesthetics ratings~\cite{harrison2015infographic}, the similarity of visualization pairs~\cite{ma2018scatternet}, ranking of visualization quality~\cite{wu2021learning}.
Relative ranking among visualizations is often collected instead of absolute ratings~\cite{ma2018scatternet}.
While natural language is an important way to interact with visualizations and communicate information and insights inside visualizations, they are usually collected in a crowdsourcing way to get more diverse and natural expressions.
Descriptions and answers on visualizations are analyzed to understand patterns in them~\cite{Kim2020, srinivasan2021collecting}.
The manual method is also used to derive templates~\cite{liu2020autocaption} or get paraphrased expressions of existing sentences~\cite{fu2020quda} to augment the collected dataset.
\revision{The manual method enables the creation of more accurate and natural datasets but costs more than automatic methods.}

\textbf{Rule-based Method.}
The raw data collected during the construction of the dataset may need further processing to be suitable for visualization tasks.
\revision{Such processing includes the rule-based method for data cleaning and data derivation.}
For instance, MultiVision~\cite{Wu2022} filtered out data tables with more than ten columns and Moreover, charts encoded over four columns from the Excel corpus in Table2Charts~\cite{Zhou2020}.
Meta information of visualizations and underlying data can be derived by rule-based methods.
For example, VizNet~\cite{viznet} is derived from meta-information with algebraic calculations.
\revision{Hulsebos et al.~\cite{sherlock} generate semantic types of attributes by matching templates.
Liu et al.~\cite{liu2020autocaption} synthesized caption templates to expand the visualization to caption dataset. }
Information regarding the data and visual mappings in visualizations is typically embedded within the visualization charts, however, such information is not always provided simultaneously.
\revision{D3 search~\cite{d3search} acquires information about underlying data and mappings by parsing the SVG-based visualizations generated by D3.}
The D3 search leverages the capability of D3 to bind data to HTML document objects, as well as heuristics of parsing visualizations, to enable data and template extraction~\cite{harper2014deconstructing}. 
The rule-based method can be applied to increase the amount of training data. For example, some icons in Visually29K~\cite{Madan2021} are generated icons by detecting blank areas and adding icons randomly and the location and size information is obtained during the generation process, as illustrated in~\autoref{fig:how_example} (2).
Natural language data like descriptions of visualization content or question-answer pairs based on charts are totally or partially generated with rule-based methods like generating based on templates in AutoCaption~\cite{liu2020autocaption}, FigureQA~\cite{figureqa}, DVQA~\cite{Kafle2018}, PlotQA~\cite{methani2020plotqa}, LEAF-QA~\cite{chaudhry2020leaf}, and STL-CQA~\cite{singh2020stl}.
These templates are manually created by the dataset creators or summarized from manually collected sentences.
FigureQA~\cite{figureqa} is further converted into visualization and caption pairs from figure caption dataset~\cite{Chen_2020_WACV}.
The rule-based method can extend datasets with a relatively small cost but it suffers from the problem of lacking diversity of real-world visualizations.

\textbf{Learning-based Method.}
Dataset augmentation can be achieved through learning-based methods. 
For instance, Beagle~\cite{beagle} trains a support vector machine to categorize visualization types based on the number and distribution of various elements.
AutoTitle~\cite{liu2023autotitle} use large language model to generate the fact to title dataset.
Tools such as DiagramFlyer~\cite{chen2015diagramflyer}, FigureSeer~\cite{siegel2016figureseer}, and the work of Choudhury et al.~\cite{ray2015architecture} extract common data from visual graphs with deep learning models like Convolutional Neural Network.
Among these works, FigureSeer~\cite{siegel2016figureseer} identifies neatly aligned text to determine the coordinates of the axes and classifies whether the text is part of the legend by using a trained random forest method.
The reverse-engineered dataset~\cite{reverse} employs Convolutional Neural Networks to classify whether each pixel is text or not, and then further categorize the labels and titles of the axes and the text in the legend based on the arrangement pattern of the text.
The inclusion of natural language elements, such as captions, questions, and answers, has become increasingly important in visualization datasets.
\removed{Learning methods can be applied to combine the benefits of the manual method and the rule-based method.}
\revision{Learning methods can also serve as supplements to existing data methods (manual-based or ruled-based), such as enhancing the diversity of natural language through large language models.}
For example, Luo et al.~\cite{luo2021synthesizing} use translation techniques in the field of natural language processing to improve the diversity and accuracy of the templates generated.
Fu et al.~\cite{fu2020quda} aims to reduce the cost of crowdsourcing annotation while collecting high-quality natural language questioning data on a larger scale. They divide the annotation process into two parts: expert users asking questions based on the task and data, and ordinary users paraphrasing the expert users' questions to increase the diversity of the natural language data.
Furthermore, Zhang et al.~\cite{zhang2022onelabeler} proposed a solution to alleviate the tedious process of manual labeling by introducing a system that utilizes active learning to generate annotations and minimize the requirement for human labor.
The learning method helps generate datasets mimicking a real-world style, however, it is important to ensure the quality of the final results as the quality of data largely influences the quality of model~\cite{vidgen2020directions}.
\remark{"however, it is important to ensure the quality of the final results" should clarify why and how}
\revision{Liu et al~\cite{liu2023autotitle} use LLM to rewrite the natural language of title content to construct a fact-to-title pair dataset. 
To ensure the quality of the training dataset, they check the quality manually.}

\section{Research Challenge and Discussion}

In recent years, \removed{the fields of visualization and machine learning have evolved, and }machine learning methods have played a crucial role in the visualization process.
Datasets serve as the foundation for these works, and many studies have recognized their importance and made initial attempts.
\revision{However, the formats used in data visualization are diverse, ranging from specifications (e.g., Vega-Lite~\cite{satyanarayan2016vegalite}), intermediate representations (e.g., D3~\cite{bostock2011d3}), vector graphics (e.g., SVG), to raster graphics (e.g., JPG).}
\removed{However, current visualization datasets suffer from issues such as a lack of standardization, small data scale, and insufficient openness.}

\subsection{Visualization Datasets Standardization}

\revision{The conversion between these visualization formats is often challenging and may involve risks such as loss of information and inability to convert with precision.}
A standardized visualization data structure should strive to be independent of specific tools and file formats and should be universally accessible to both machines and humans.
This pursuit aims to establish a unified data structure for expressing visual information that can become a new standard for visualization files.
Such a data structure can inherently encode the maximum amount of visual data and mapping information.

Another kind of standardization for visualization datasets is the use of a multi-modal transformer.
Multi-modal universal methods for different visualization learning tasks, such as classification, visualization representation, and visualization-natural language representation, should be developed by establishing a shared pre-trained neural network foundation.
\revision{In natural language and images domains, multi-modal universal models (for example, CLIP~\cite{radford2021learning} and ViT~\cite{dosovitskiy2020image}) seek to represent different forms of information through a unified latent space, thus supporting translation between these forms.}
By leveraging the large-scale, weakly-supervised visualization-natural language correspondences in network media, models similar to those used in large-scale image-natural language learning can be trained using large-scale, weakly-supervised datasets to learn universal values~\cite{mahajan2018exploring}.
\remark{Add a reference.}

\begin{figure}[htb]
	\centering
    \begin{minipage}[b]{0.5\textwidth}
        \includegraphics[width=0.99\columnwidth]{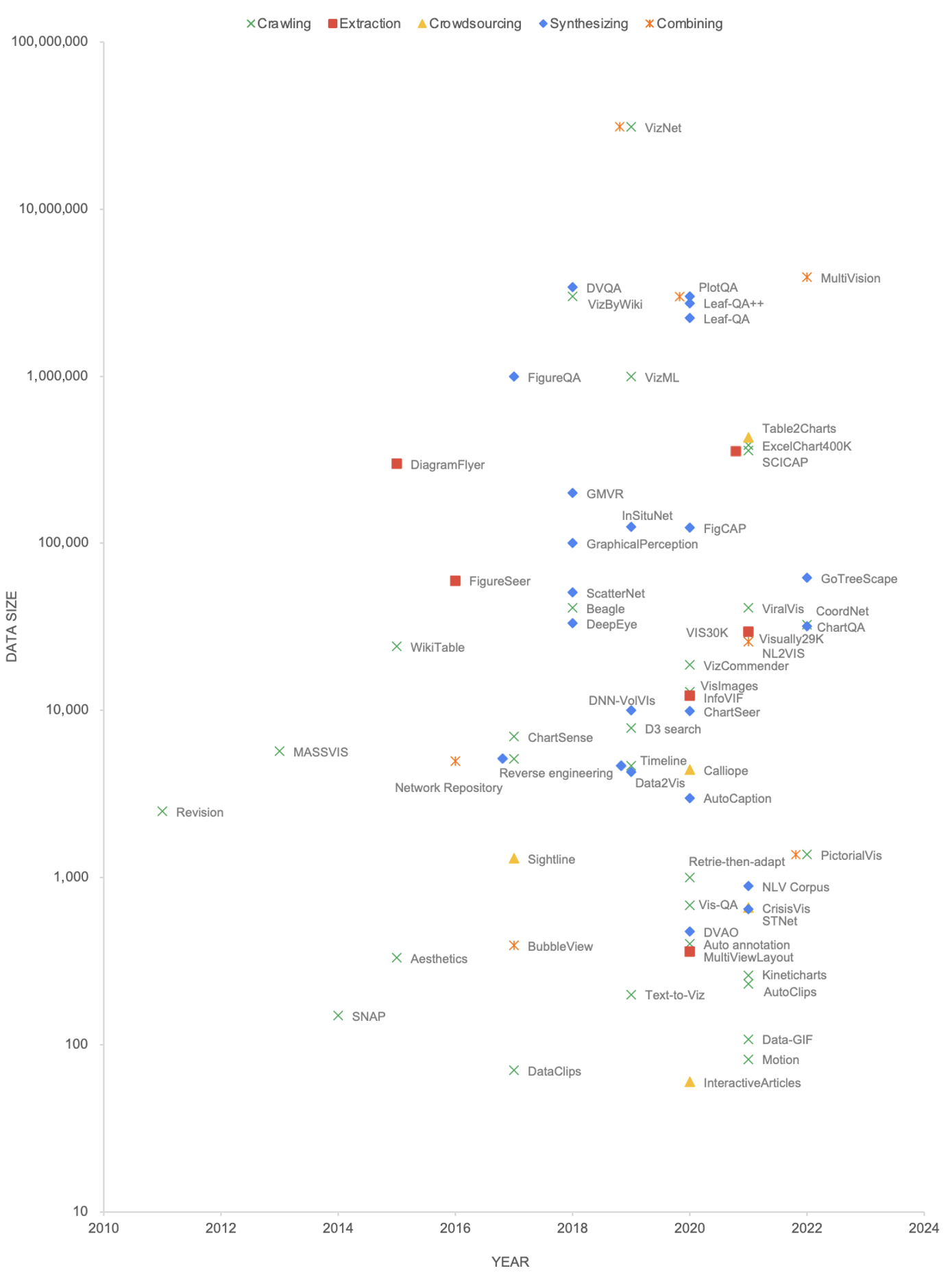}
    \end{minipage}
    
	\caption{\label{fig:dataset_scatterplot} The x-axis encodes the release year, the y-axis presents the number of data items, and the color represents the data construction method.
    \revision{The scale of datasets used for visualization is growing over time, and as these datasets begin to expand, existing visualization datasets are often limited in terms of scale. Consequently, large-scale datasets are typically constructed through methods such as web crawling or synthesizing.}
    \removed{We can see that existing visualization datasets are often limited in scale and large-scale datasets are always constructed by crawling or synthesizing.}}
\end{figure}

\subsection{Visualization Datasets Size}

The size, construction method, and release year of the datasets are shown in ~\autoref{fig:dataset_scatterplot}.
The figure demonstrates an evident trend in recent years: a significant increase in the number of larger and more extensive visualization datasets.
In terms of construction methods, there has been a gradual shift from crawling methods to the inclusion of extraction, crowdsourcing, synthesizing, and combining techniques.
\revision{Compared with datasets in image\footnote{Image: \url{https://paperswithcode.com/datasets?mod=images}} or natural language area\footnote{Text: \url{https://paperswithcode.com/datasets?mod=texts}}, visualization datasets are often limited in scale because of heavy cost in manual annotation, and large-scale data is typically obtained through crawling or rule-based synthesis without human labeling.}
However, many machine learning tasks, particularly end-to-end tasks, require a large amount of labeled training data. Data labeling is a topic of interest in academic circles, and studies such as Zhang et al.~\cite{zhang2022onelabeler} summarize the primary processes involved in labeling.
To label large-scale data, a significant amount of manual effort is typically required. Therefore, it is worthwhile to investigate ways to expand the scale of data while reducing manpower investment and ensuring quality.
One possible approach is to establish an intelligent data annotation process that involves human input to ensure the quality and quantity of visualization annotations.
This process requires the machine to continuously learn from the annotation results and provide feedback to the annotation loop, as MI3 did~\cite{zhang2021mi3}.
As users become more familiar with the annotation process, the machine can provide more initial annotations, enabling users to focus on the data sets that the machine does not understand and avoiding an infinite annotation process.

\subsection{Visualization Datasets Openness}
Easy access to visualization datasets can be facilitated through three primary methods. The most commonly used approach is to provide users with a download link, where the entire dataset is compressed and made available for local usage after downloading. This method is best suited for datasets with a relatively low volume. However, for large datasets, this method can be time-consuming, which presents a significant constraint to the widespread utilization and dissemination of these datasets.
An alternative approach is to provide users with data construction techniques. For synthesized datasets, the authors should provide a detailed description of the construction methodology, including the models, coefficients, initial parameters, simulation equations, and other relevant details. By providing this information, other researchers can independently reproduce the synthetic datasets locally, eliminating the need for network transmission of the datasets and improving their usage. However, this method is not appropriate for non-synthetic datasets, particularly those that require collection and annotation processes, as these processes contribute significant value to the release of the datasets.

Cloud services have emerged as a prevalent method for data management and access. These services involve hosting datasets on servers that respond to client requests by transmitting the data. Many of these servers are located on supercomputers, which offer additional benefits such as online data processing capabilities. This eliminates the need for clients to download datasets, as they can simply log into their accounts to access them.
\removed{Recently, remote data access through cloud services has gained popularity due to its versatility and avoidance of data transmission issues.}
Some datasets, such as Gaia\cite{moitinho_gaia_2017}, are actively seeking to establish a platform community that not only facilitates remote data access but also provides additional resources, such as user-friendly data processing methods and shared discussion spaces. While building such a platform is a significant undertaking, it has the potential to create a more supportive research and communication environment for the community.

Despite the significant progress in the field of visualization and machine learning, there is currently a lack of a unified and accessible platform for sharing visualization datasets. In the natural language and image domains, standard datasets are available for most tasks, including both training and test sets. Relevant papers for a specific task are also listed for comparison and evaluation. In contrast, AI-for-visualization approaches often spend a significant amount of time on data construction, and it is difficult to conduct cross-comparative evaluations due to data openness issues.
Therefore, we call for a more open data platform that provides data standards, supports data uploads, and provides standard training and benchmark data for visualization tasks. Such a platform would help to streamline the data construction process and promote better cross-comparative evaluations of visualization approaches.

\ifCLASSOPTIONcompsoc
  \section*{Acknowledgments}
\else
  \section*{Acknowledgment}
\fi
This work is supported by NSFC No. 62272012. 
\ifCLASSOPTIONcaptionsoff
  \newpage
\fi

\bibliographystyle{IEEEtran}
\bibliography{manuscript}

\end{CJK*}
\end{document}